\documentclass[%
 reprint,
 superscriptaddress,
nofootinbib,
 amsmath,amssymb,
 aps,
]{revtex4-1}

\usepackage{threeparttable}
\usepackage{braket}
\usepackage{graphicx}
\usepackage{dcolumn}
\usepackage{bm}
\usepackage{hyperref}
\usepackage{cleveref}
\usepackage{xcolor}

\begin{document}

\preprint{APS/123-QED}

\title{Systematics of $E2$ strength in the sd-shell with the valence-space in-medium similarity renormalization group}

\author{S.~R.~Stroberg}
\email{sstroberg@anl.gov}
\affiliation{TRIUMF, Vancouver, BC V6T 2A3, Canada}
\affiliation{Department of Physics, University of Washington, Seattle WA, USA}
\affiliation{Physics Division, Argonne National Laboratory, Lemont IL, USA}
\author{J.~Henderson}
\email{jack.henderson@surrey.ac.uk}
\affiliation{TRIUMF, Vancouver, BC V6T 2A3, Canada}
\affiliation{Department of Physics, University of Surrey, Guildford, GU2 7XH, United Kingdom}
\affiliation{Lawrence Livermore National Laboratory, Livermore, CA 94550, USA}
\author{G.~Hackman}
\affiliation{TRIUMF, Vancouver, BC V6T 2A3, Canada}
\author{P.~Ruotsalainen}
\affiliation{University of Jyv\"askyl\"a, Department of Physics, P. O. Box 35, FI-40014 University of Jyv\"askyl\"a, Finland}
\author{G.~Hagen}
\affiliation{Physics Division, Oak Ridge National Laboratory, Oak Ridge, Tennessee 37831, USA}
\affiliation{Department of Physics and Astronomy, University of Tennessee, Knoxville, Tennessee 37996, USA}
\affiliation{TRIUMF, Vancouver, BC V6T 2A3, Canada}
\author{J.~D.~Holt}
\affiliation{TRIUMF, Vancouver, BC V6T 2A3, Canada}
\affiliation{Department of Physics, McGill University, 3600 Rue University, Montr\'eal, QC H3A 2T8, Canada}

\date{\today}

\begin{abstract}
\begin{description}
\item[Background] Recent developments in {\it ab initio} nuclear theory demonstrate promising results in medium- to heavy-mass nuclei. A particular challenge for many of the many-body methodologies, however, is an accurate treatment of the electric-quadrupole, $E2$, strength associated with collectivity.
\item[Purpose] The valence-space in-medium similarity renormalization group (VS-IMSRG) is a particularly powerful method for accessing medium- and high-mass nuclei but has been found to underpredict $E2$ strengths. The purpose of this work is to evaluate the isospin dependence of this underprediction.
\item[Methods] We perform a systematic comparison of valence-space in-medium similarity renormalization group (VS-IMSRG) calculations with available literature. We make use of isoscalar and isovector contributions to the $E2$ matrix elements to assess isoscalar and isovector contributions to the missing strength.
\item[Results] It is found that the $E2$ strength is consistent throughout $T_z=\left|\frac{1}{2}\right|$, $T_z=\left|1\right|$, $T_z=\left|\frac{3}{2}\right|$ and $T_z=2$ pairs within the $sd$-shell. Furthermore, no isovector contribution to the deficiency is identified.
\item[Conclusions] A comparison with toy-models and coupled-cluster calculations is used to discuss potential origins of the missing strength, which arises from missing many-particle, many-hole excitations out of the model space. The absence of any significant isovector contribution to the missing $E2$ strength indicates that the $E2$ strength discrepancy, and therefore any correction, is largely independent of the isospin of the nuclei in question.
\end{description}
\end{abstract}

\pacs{Valid PACS appear here}
\maketitle

\section{Introduction}

Modelling the atomic nucleus for all but the lightest systems requires an approximate many-body method in order to make the problem computationally tractable. The approximation used will naturally result in some deficiencies in the reproduction of observable properties. Understanding how these deficiencies behave and can be corrected for is essential in ensuring no loss of predictive power arising from the choice of many-body method. This is especially true as a model is used to probe unexplored regions of the nuclear landscape where experimental data is completely absent. The region about the line of $N$$=$$Z$ therefore provides an exceptional testing ground for models, allowing for any isospin-dependence of model deficiencies to be identified through careful comparison with high-precision data. This gives confidence that at extreme isospin (i.e. approaching the proton and neutron drip line), corrections for deficiencies in the many-body method can still reliably be employed. 

The strength of electric quadrupole ($E2$) transitions between excited states in nuclei is closely related to the collective motion of nucleons arising from deviations of the nucleus from sphericity. The theoretical reproduction of this $E2$ strength has long been problematic for, for example, the phenomenological nuclear shell model. The shell model begins with a spherical potential, and collectivity has therefore to be introduced through multi-particle multi-hole (mp-mh) excitations, with large contributions from configurations outside the shell-model, or valence, space. In a recent work~\cite{ref:Henderson_18}, we investigated the ability of modern, microscopically derived nuclear theory to reproduce $E2$ strength without the use of adjustments to the nucleon charges ({\it effective charges}) that are required in shell-model methodologies. It was found that the symplectic no-core shell model (NCSpM)~\cite{ref:Tobin_14} reproduced the experimental data well using an appropriate subset of mp-mh excitations to account for the collective motion of the nucleons. {\it Ab initio} valence-space in-medium similarity renormalization group (VS-IMSRG) calculations meanwhile failed to reproduce the absolute $E2$ strengths in Ref.~\cite{ref:Henderson_18} but provided superior qualitative description of the experimental evolution of $E2$ strength in the $sd$-shell, as compared to phenomenological shell-model calculations. 

In low-mass systems, where quasi-exact methodologies such as the no-core shell model (NCSM) can be employed, effective charges can be reliably calculated and indeed studied~\cite{ref:Navratil_97,ref:Stetcu_05,ref:Lisetskiy_09}. Beyond this region however, some approximation must be employed. While the use of symmetry adjusted methods such as the NCSpM and the closely-related symmetry-adjusted no-core shell model allow for calculations beyond the limits of, for example, the NCSM, they remain tractable only for nuclei below about mass 40. The VS-IMSRG methodology, however, is far more widely applicable, with calculations approaching convergence even in the vicinity of \textsuperscript{132}Sn~\cite{ref:Miyagi_22} and beyond~\cite{ref:Hu_22}, and spanning both proton and neutron driplines~\cite{ref:Stroberg_21}. It is therefore imperative that the model deficiencies (and corrections thereof) are well understood, especially with regards to the isospin degree of freedom as future studies move into heavier, more neutron-rich regions of the nuclear landscape.

In this work, we perform a systematic comparison between experimental $E2$ strengths and both shell model and VS-IMSRG calculations. By making use of the isoscalar and isovector matrix element formalism of Brown et al.~\cite{ref:Brown_82}, we are able to assess not only the $E2$ strength deficiency with a wider selection of data, but also to isolate any isovector terms that might be introduced in the many-body approximation. 

\section{Calculations and Data}

$B(E2)$ values extracted in the recent work are presented in Table~\ref{tab:T_1_2}, along with the present state of knowledge for $T_z=\pm\frac{1}{2}$ nuclei with $19\leq A\leq31$. Also shown are the results of calculations using the a \emph{ab initio} VS-IMSRG methodology~\cite{ref:Tsukiyama_12,ref:Bogner_14,ref:Stroberg_17,ref:Stroberg_19}, with a consistently evolved valence-space $E2$ operator~\cite{ref:Parzuchowski_17} which does not rely on the use of any effective charges. 
The VS-IMSRG calculations were performed using the EM1.8/2.0 interaction~\cite{ref:Hebeler_11,ref:Simonis_16}, which was generated by SRG evolution~\cite{ref:Bogner_07} of the chiral N$^3$LO NN interaction of Entem and Machleidt~\cite{ref:Entem_03}, and adding a non-locally regulated N$^2$LO 3N interaction with the low energy constants adjusted to reproduce the triton binding energy and the $^{4}$He matter radius.
While only constrained with $A\leq 4$ data, this interaction gives a remarkable reproduction of binding energies and spectroscopy up to at least $A\approx 100$, with a general underprediction of radii~\cite{ref:Simonis_17,ref:Morris_18,ref:Stroberg_21}.
Calculations are performed in a harmonic oscillator basis of $\hbar\omega=20$~MeV with $2n+\ell\leq e_{max}$=12
and an additional truncation on the three body matrix elements $e_1+e_2+e_3\leq E_{3max}$=16.
Following a Hartree-Fock calculation, all operators are truncated at the normal-ordered two-body level. Also shown are shell-model calculations performed with the USDB interaction~\cite{ref:USDB}.
In both cases, shell model diagonalizations are performed using the code NuShellX~\cite{ref:NushellX}, and the transition densities needed for the VS-IMSRG operators are computed using the nutbar code~\cite{ref:nutbar}. The nominal USDB effective charges of $e_p = 1.36$ and $e_n = 0.45$ were used for all phenomenological shell-model calculations. Shown in Table~\ref{tab:T_1} are $T_z=\pm1$ nuclei, in Table~\ref{tab:T_3_2} are $T_z=\pm\frac{3}{2}$ nuclei and in Table~\ref{tab:T_2} $T_z=\pm2$ nuclei. 

\section{Discussion}

\begin{figure}
\centerline{\includegraphics[width=\linewidth]{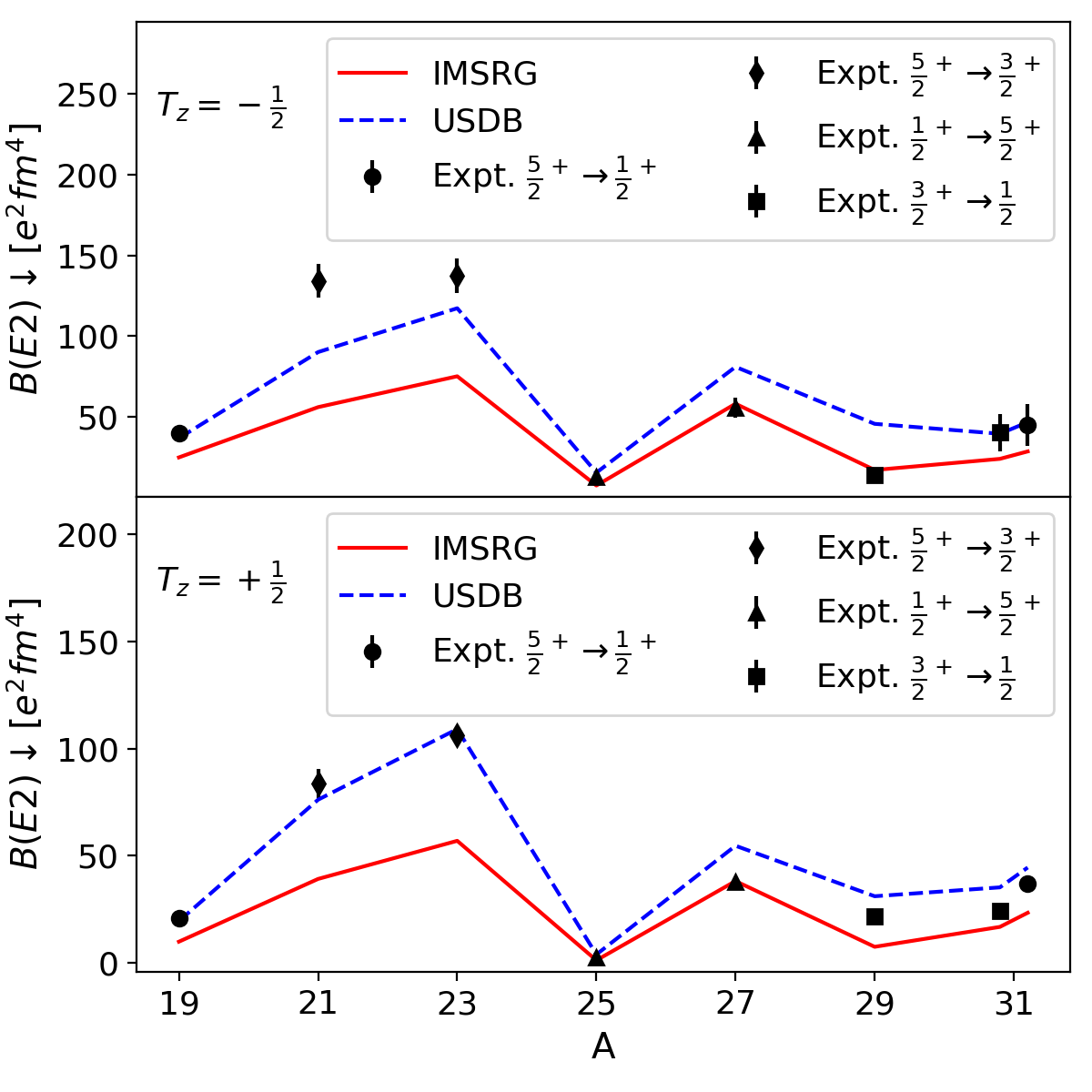}}
\caption{$B(E2)\downarrow$ values for (top) $T_z=-\frac{1}{2}$ and (bottom) $T_z=+\frac{1}{2}$ nuclei. See text for details of the theoretical VS-IMSRG and USDB calculations.}
\label{fig:BE2_Tz_1_2}
\end{figure}

\begin{figure}
\centerline{\includegraphics[width=\linewidth]{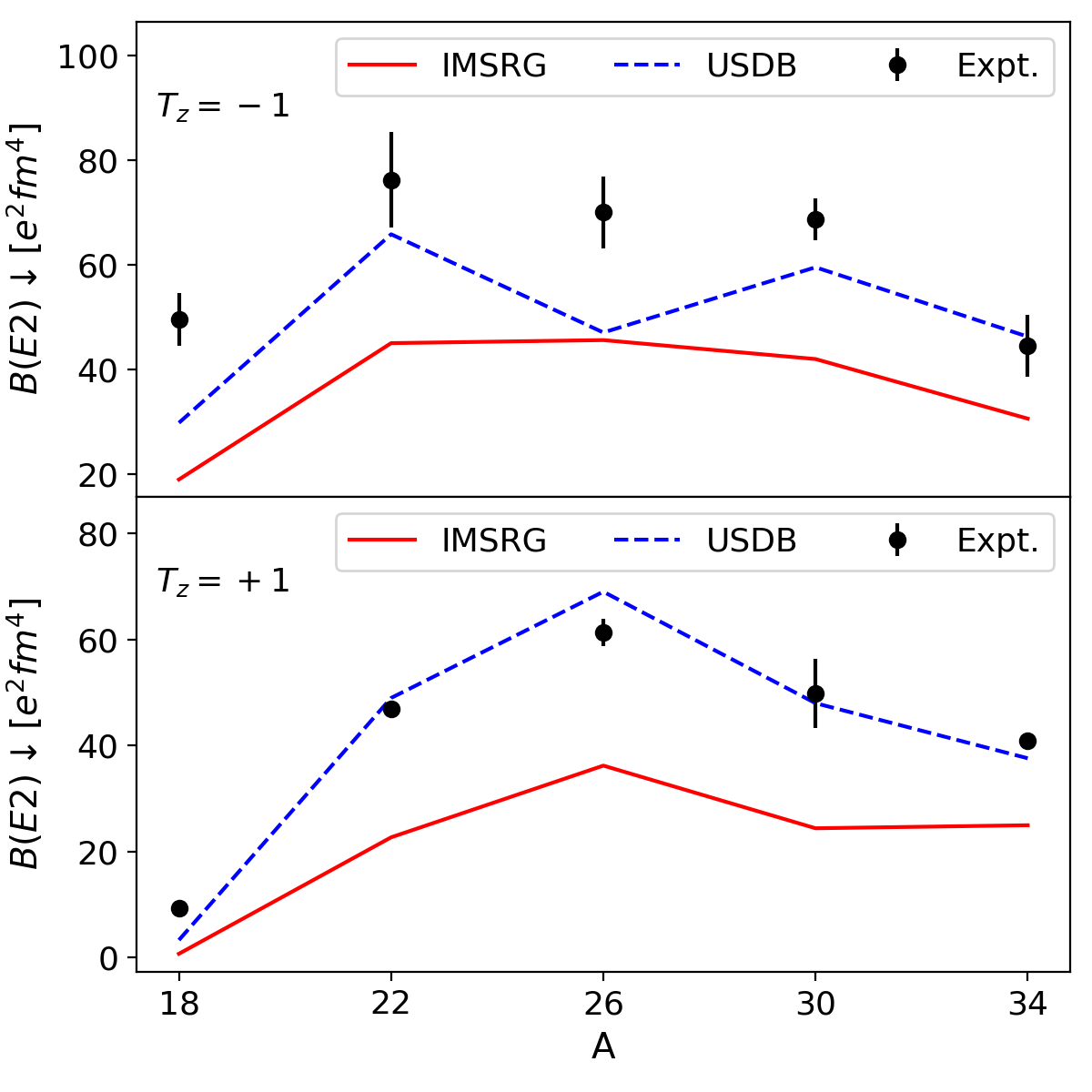}}
\caption{$B(E2)\downarrow$ values for (top) $T_z=-1$ and (bottom) $T_z=+1$ nuclei. See text for details of the theoretical VS-IMSRG and USDB calculations.}
\label{fig:BE2_Tz_1}
\end{figure}

\begin{figure}
\centerline{\includegraphics[width=\linewidth]{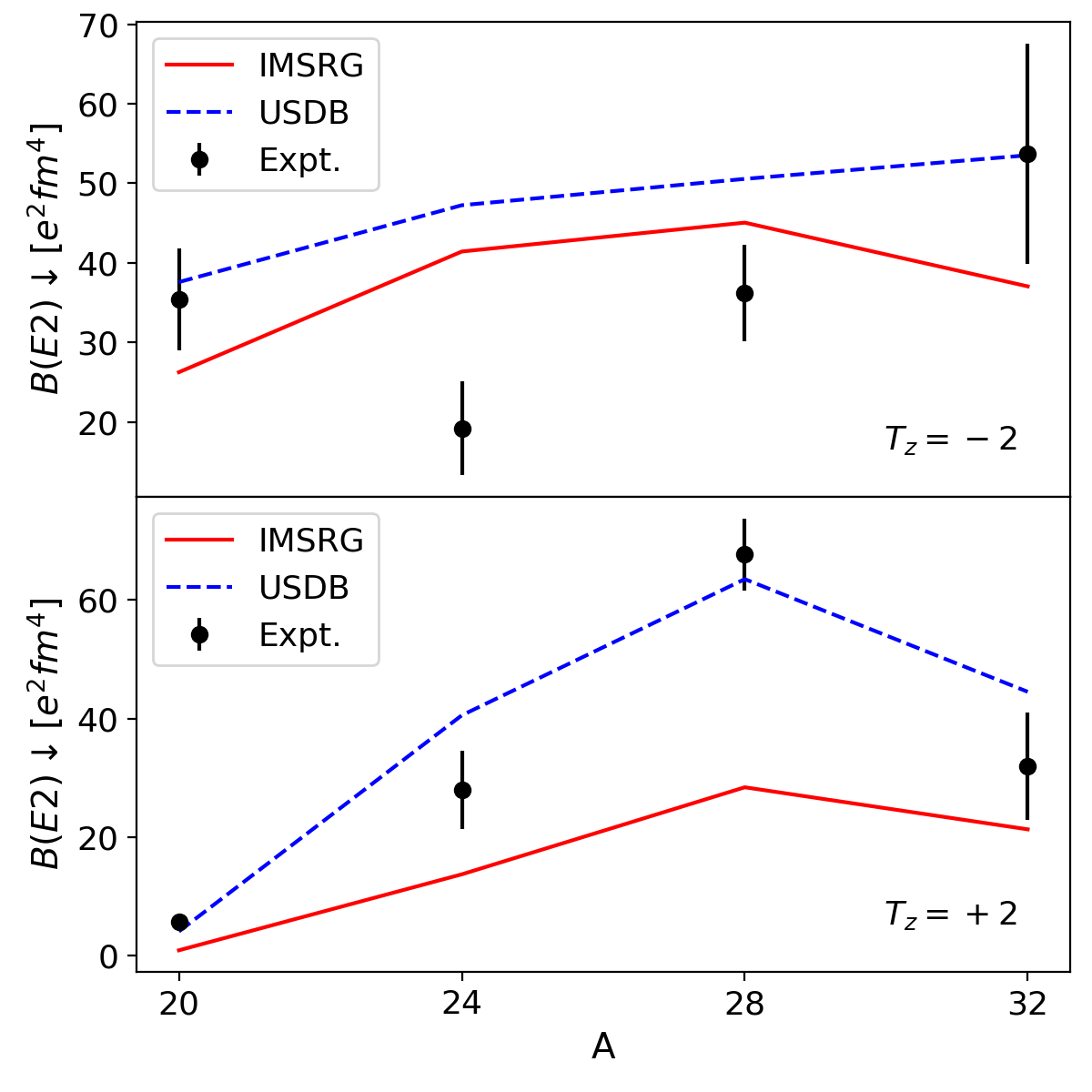}}
\caption{$B(E2)\downarrow$ values for (top) $T_z=-2$ and (bottom) $T_z=+2$ nuclei. See text for details of the theoretical VS-IMSRG and USDB calculations.}
\label{fig:BE2_Tz_2}
\end{figure}

The experimental and calculated values for $\left|T_z\right| = \frac{1}{2}$,  $\left|T_z\right| = 1$ and $\left|T_z\right| = 2$ are shown in Fig~\ref{fig:BE2_Tz_1_2}, Fig~\ref{fig:BE2_Tz_1} and Fig~\ref{fig:BE2_Tz_2}, respectively. Due to the limited available experimental data, values for $\left|T_z\right| = \frac{3}{2}$ are not plotted here, but a similar plot can be found in Ref.~\cite{ref:Ruotsalainen_19}. Clearly, from the results presented in Tables.~\ref{tab:T_1_2} to~\ref{tab:T_2} and Figures~\ref{fig:BE2_Tz_1_2} to~\ref{fig:BE2_Tz_2}, the VS-IMSRG calculations underpredict absolute $B(E2)$ strength, as was previously reported~\cite{ref:Henderson_18, ref:Ruotsalainen_19}. As would be expected, the USDB calculations with nominal effective charges reproduce the absolute strength relatively well. 

\begin{table}
\begin{threeparttable}
\caption{$B(E2)$ values for transitions in $\left|T_z=\frac{1}{2}\right|$ nuclei in the $sd$ shell, comparing experimental values with those calculated using VS-IMSRG with the EM1.8/2.0 interaction and with the USDB shell-model interaction. Shell-model calculations used effective charges of $e_p = 1.36$ and $e_n = 0.45$.}
\label{tab:T_1_2}
\begin{ruledtabular}
\begin{tabular}{lccccccc}
 & & &  \multicolumn{3}{c}{B(E2)$\downarrow$ [e$^2$fm$^4$]} & \\
 \hline \\[-7pt]
Isotope & $J^\pi_i$ & $J^\pi_f$ & Expt. & VS-IMSRG & USDB & Ref. (Expt.) \\[+2pt]
 \hline \\[-8pt]
$^{19}$Ne 	& $\frac{5}{2}^+_1$ 	& $\frac{1}{2}^+_1$ 	& 39.8 (15) 	& 25.0 	& 36.9	&	\cite{ref:ENSDF} 	\\[+1pt]
$^{19}$F 		& $\frac{5}{2}^+_1$ 	& $\frac{1}{2}^+_1$ 	& 20.9 (2)	 	& 9.8 	& 19.4	& 	\cite{ref:ENSDF} 	\\[+1pt]
$^{21}$Na 	& $\frac{5}{2}^+_1$ 	& $\frac{3}{2}^+_1$ 	& 136.5 (92) 	& 56.1 	& 90.2	&	\cite{ref:ENSDF} 	\\[+1pt]
$^{21}$Ne		& $\frac{5}{2}^+_1$ 	& $\frac{3}{2}^+_1$ 	& 87.5 (58)	& 39.1 	& 76.1	& 	\cite{ref:ENSDF} 	\\[+1pt]
$^{23}$Mg 	& $\frac{5}{2}^+_1$ 	& $\frac{3}{2}^+_1$ 	& 135 (15) 	& 75.2 	& 117.3	&	\cite{ref:Henderson_21}		\\[+1pt]
$^{23}$Na		& $\frac{5}{2}^+_1$ 	& $\frac{3}{2}^+_1$ 	& 106 (4) 	& 56.9 	& 109.1 	&	\cite{ref:Henderson_21}			\\[+1pt]
$^{25}$Al 		& $\frac{1}{2}^+_1$ 	& $\frac{5}{2}^+_1$ 	& 13.2 (3) 		& 7.6 	& 3.8 	&	\cite{ref:ENSDF} 	\\[+1pt]
$^{25}$Mg	& $\frac{1}{2}^+_1$ 	& $\frac{5}{2}^+_1$ 	& 2.44 (4)		& 1.09 	& 3.03 	&	\cite{ref:ENSDF} 	\\[+1pt]
$^{27}$Si 		& $\frac{1}{2}^+_1$ 	& $\frac{5}{2}^+_1$ 	& 55.7 (64)	& 58.2 	& 81.0 	&	\cite{ref:ENSDF} 	\\[+1pt]
$^{27}$Al		& $\frac{1}{2}^+_1$ 	& $\frac{5}{2}^+_1$ 	& 37.8 (11) 	& 38.1 	& 54.6 	&	\cite{ref:ENSDF} 	\\[+1pt]
$^{29}$P 		& $\frac{3}{2}^+_1$ 	& $\frac{1}{2}^+_1$ 	& 14.3 (27) 	& 17.2 	& 45.8 	&	\cite{ref:ENSDF} 	\\[+1pt]
$^{29}$Si		& $\frac{3}{2}^+_1$ 	& $\frac{1}{2}^+_1$ 	& 21.7 (21)	& 7.4 	& 31.0 	&	\cite{ref:ENSDF} 	\\[+1pt]
$^{31}$S 		& $\frac{3}{2}^+_1$ 	& $\frac{1}{2}^+_1$ 	& 40.5 (116) 	& 24.1 	& 39.6 	&	\cite{ref:ENSDF} 	\\[+1pt]
$^{31}$P		& $\frac{3}{2}^+_1$ 	& $\frac{1}{2}^+_1$ 	& 24.3 (35)	& 16.7 	& 35.2 	&	\cite{ref:ENSDF} 	\\[+1pt]
$^{31}$S 		& $\frac{5}{2}^+_1$ 	& $\frac{1}{2}^+_1$ 	& 45.1 (127) 	& 28.7 	& 46.8 	&	\cite{ref:ENSDF} 	\\[+1pt]
$^{31}$P		& $\frac{5}{2}^+_1$ 	& $\frac{1}{2}^+_1$ 	& 37.0 (29)	& 23.3 	& 44.4 	&	\cite{ref:ENSDF} 	\\[+1pt]
\end{tabular}
\end{ruledtabular}
\vspace{-10pt}
\end{threeparttable}
\end{table}

\begin{table}
\begin{threeparttable}
\caption{As Table~\ref{tab:T_1_2} but for $\left|T_z=1\right|$ nuclei.}
\label{tab:T_1}
\begin{ruledtabular}
\begin{tabular}{lccccccc}
 & \multicolumn{3}{c}{B(E2)$\downarrow$ [e$^2$fm$^4$]} & \\
 \hline \\[-7pt]
Isotope 		& 	Expt. 	& 	VS-IMSRG	&	USDB 	& Ref. (Expt.) 	\\[+2pt]
 \hline \\[-8pt]
$^{18}$Ne 	&	49.6 (50)	& 	19.0		& 	29.8		& \cite{ref:ENSDF}							\\[+1pt]
$^{18}$O 		& 	9.3 (3)	& 	0.7		& 	3.3		& \cite{ref:ENSDF}							\\[+1pt]
$^{22}$Mg 	& 	76.2 (92)	& 	45.0		& 	65.8		& \cite{ref:Henderson_18}						\\[+1pt]
$^{22}$Ne		& 	46.9 (5)	& 	22.7		& 	49.0		& \cite{ref:Henderson_18, ref:ENSDF}$^\dagger$	\\[+1pt]
$^{26}$Si 		& 	70.0 (69)	& 	45.6		& 	47.1		& \cite{ref:ENSDF}							\\[+1pt]
$^{26}$Mg	& 	61.3 (26)	& 	36.2		& 	69.0		& \cite{ref:ENSDF}							\\[+1pt]
$^{30}$S 		& 	68.7 (40)	& 	42.0		& 	59.5		& \cite{ref:ENSDF}							\\[+1pt]
& 	43.7 (44)	&  		& 		& \cite{ref:Petkov_17}$^*$							\\[+1pt]
$^{30}$Si		& 	49.9 (65)	& 	24.4		& 	48.0		& \cite{ref:ENSDF}							\\[+1pt]
$^{34}$Ar 		& 	44.5 (59)	& 	30.6		& 	46.3		& \cite{ref:ENSDF}							\\[+1pt]
$^{34}$S		& 	40.8 (11)	& 	24.9		& 	37.6		& \cite{ref:ENSDF}							\\[+1pt]
\end{tabular}
\end{ruledtabular}
\begin{tablenotes}
\item[$\dagger$] - Weighted average of values in Ref.~\cite{ref:Henderson_18} and Ref.~\cite{ref:ENSDF}
\item[$^*$] Two experimental values for \textsuperscript{30}S disagree significantly. The analysis in the present work uses the evaluated value from Ref.~\cite{ref:ENSDF}.
\end{tablenotes}
\vspace{-10pt}
\end{threeparttable}
\end{table}

\begin{table}
\begin{threeparttable}
\caption{As Table~\ref{tab:T_1_2} but for $\left|T_z=\frac{3}{2}\right|$ nuclei.}
\label{tab:T_3_2}
\begin{ruledtabular}
\begin{tabular}{lccccccc}
 & & &  \multicolumn{3}{c}{B(E2)$\downarrow$ [e$^2$fm$^4$]} & \\
 \hline \\[-7pt]
Isotope 		& 	$J^\pi_i$ 			& 	$J^\pi_f$ 			& 	Expt. 		& 	VS-IMSRG 	& 	USDB 	& 	Ref. (Expt.) 			\\[+2pt]
 \hline \\[-8pt]
$^{21}$Mg 	& 	$\frac{1}{2}^+_1$	&	$\frac{5}{2}^+_1$	&	131.1 (14) 	& 	94.6		& 	132.0	&	\cite{ref:Ruotsalainen_19}	\\[+1pt]
$^{21}$F 		& 	$\frac{1}{2}^+_1$	&	$\frac{5}{2}^+_1$	& 	54.0 ( 55)		& 	23.6		& 	60.5		&	\cite{ref:ENSDF} 	\\[+1pt]
$^{21}$Mg 	& 	$\frac{9}{2}^+_1$	&	$\frac{5}{2}^+_1$	& 	83.7 (140)		& 	21.3		& 	55.6		&	\cite{ref:Ruotsalainen_19}	\\[+1pt]
$^{21}$F 		& 	$\frac{9}{2}^+_1$	&	$\frac{5}{2}^+_1$	& 	14.3 (21)		& 	6.4		& 	16.8		&	\cite{ref:VonMoss_15} 	\\[+1pt]
$^{33}$Ar 		&	$\frac{3}{2}^+_1$	&	$\frac{1}{2}^+_1$	& 	40.2	(94)		& 	19.8		& 	33.6		&	\cite{ref:Wendt_14} 		\\[+1pt]
$^{33}$P		&	$\frac{3}{2}^+_1$	&	$\frac{1}{2}^+_1$	& 	62.9 (252)		& 	18.6		& 	39.7		&	\cite{ref:ENSDF} 		\\[+1pt]
$^{33}$Ar 		& 	$\frac{5}{2}^+_1$	&	$\frac{1}{2}^+_1$	& 	36.5 (101)		& 	25.4		& 	45.4		&	\cite{ref:ENSDF} 		\\[+1pt]
$^{33}$P		&	$\frac{5}{2}^+_1$	&	$\frac{1}{2}^+_1$	& 	32.1 (50)		& 	13.6		& 	31.5		&	\cite{ref:ENSDF} 		\\[+1pt]
\end{tabular}
\end{ruledtabular}
\vspace{-10pt}
\end{threeparttable}
\end{table}

\begin{table}
\begin{threeparttable}
\caption{As Table~\ref{tab:T_1_2} but for $\left|T_z=2\right|$ nuclei.}
\label{tab:T_2}
\begin{ruledtabular}
\begin{tabular}{lccccccc}
 & \multicolumn{3}{c}{B(E2)$\downarrow$ [e$^2$fm$^4$]} & \\
 \hline \\[-7pt]
Isotope 		& 	Expt. 		& 	VS-IMSRG 	& 	USDB 	& 	Ref. (Expt.) \\[+2pt]
 \hline \\[-8pt]
$^{20}$Mg 	& 	35.4 (64)		& 	26.3		&	37.6		&	\cite{ref:ENSDF} 		\\[+1pt]
$^{20}$O		& 	5.8 (2)		& 	0.9		&	4.1		&	\cite{ref:ENSDF} 		\\[+1pt]
$^{24}$Si 		& 	19.1	(59)		& 	41.4		&	47.3		&	\cite{ref:ENSDF} 		\\[+1pt]
$^{24}$Ne		& 	28.0	(66)		& 	13.8		&	40.6		&	\cite{ref:ENSDF} 		\\[+1pt]
$^{28}$S 		& 	36.2	(60) 		& 	45.1		&	50.6		&	\cite{ref:ENSDF} 		\\[+1pt]
$^{28}$Mg	& 	67.7	(61)		& 	28.4		&	63.5		&	\cite{ref:ENSDF} 		\\[+1pt]
$^{32}$Ar 		& 	53.7	(139)		& 	37.1		&	53.5		&	\cite{ref:ENSDF} 		\\[+1pt]
$^{32}$Si		& 	32.0	(91)		& 	21.3		&	44.5		&	\cite{ref:ENSDF} 		\\[+1pt]
\end{tabular}
\end{ruledtabular}
\vspace{-10pt}
\end{threeparttable}
\end{table}

In order to understand the nature of the missing strength in the VS-IMSRG calculations, it bears briefly summarizing the many-body method. An approximately unitary transformation is performed on the Hamiltonian in the large Hilbert space (here 13 
major oscillator shells) so as to decouple a tractable valence space - analogous to a shell-model space - from the full space. The result is an effective Hamiltonian in which couplings to excitations out of the valence space  are suppressed.
The use of a unitary transformation means that - in principle - no physics is lost in this process. One can then perform configuration interaction calculations within the decoupled model space and capture all physics, even that which might have involved couplings between the core and the external model-space in the original Hamiltonian. The appeal of this method is clear, as it provides a computationally tractable valence-space Hamiltonian without losing physics information from the larger space. It is also necessary to consistently evolve all operators for use in the transformed model-space. This evolution inevitably induces three- and higher-body components, which in practice must be truncated to make the problem tractable. The operator evolution is therefore truncated at the normal-ordered two-body level---the IMSRG(2) approximation---resulting in an inevitable loss of information.

Any disagreement with experiment will be due to one of two sources: deficiencies in the input Hamiltonian (e.g. truncation of the EFT expansion), and truncations in the many-body solution. It is likely that a large fraction of the missing strength in the VS-IMSRG calculations arises from the many-body side (though there will at least be some impact from the small radii). We might expect, however, that the inclusion of two-body elements in the operator evolution will suppress the isovector component of this deficiency. The first-order contribution from out-of-space effects couples an in-space nucleon, with an out-of-space nucleon. Due to the dominance of the $T=0$ channel over $T=1$ in the interaction, the coupling is strongest between $np$ pairs. Since the $E2$ operator couples to the charge of the nucleon, this first-order correction might be expected to predominantly affect out-of-space protons coupling to in-space neutrons, as has been discussed in terms of perturbation theory~\cite{ref:Siegel_70}. These effects are captured by the IMSRG(2) approximation. For higher orders, the out-of-space couplings become more numerous, necessarily involving multiple nucleons and more configurations, meaning any individual isovector coupling contributes proportionally less to the ensemble of configurations and yielding an approximately isoscalar net effect.

\begin{figure}
\centerline{\includegraphics[width=\linewidth]{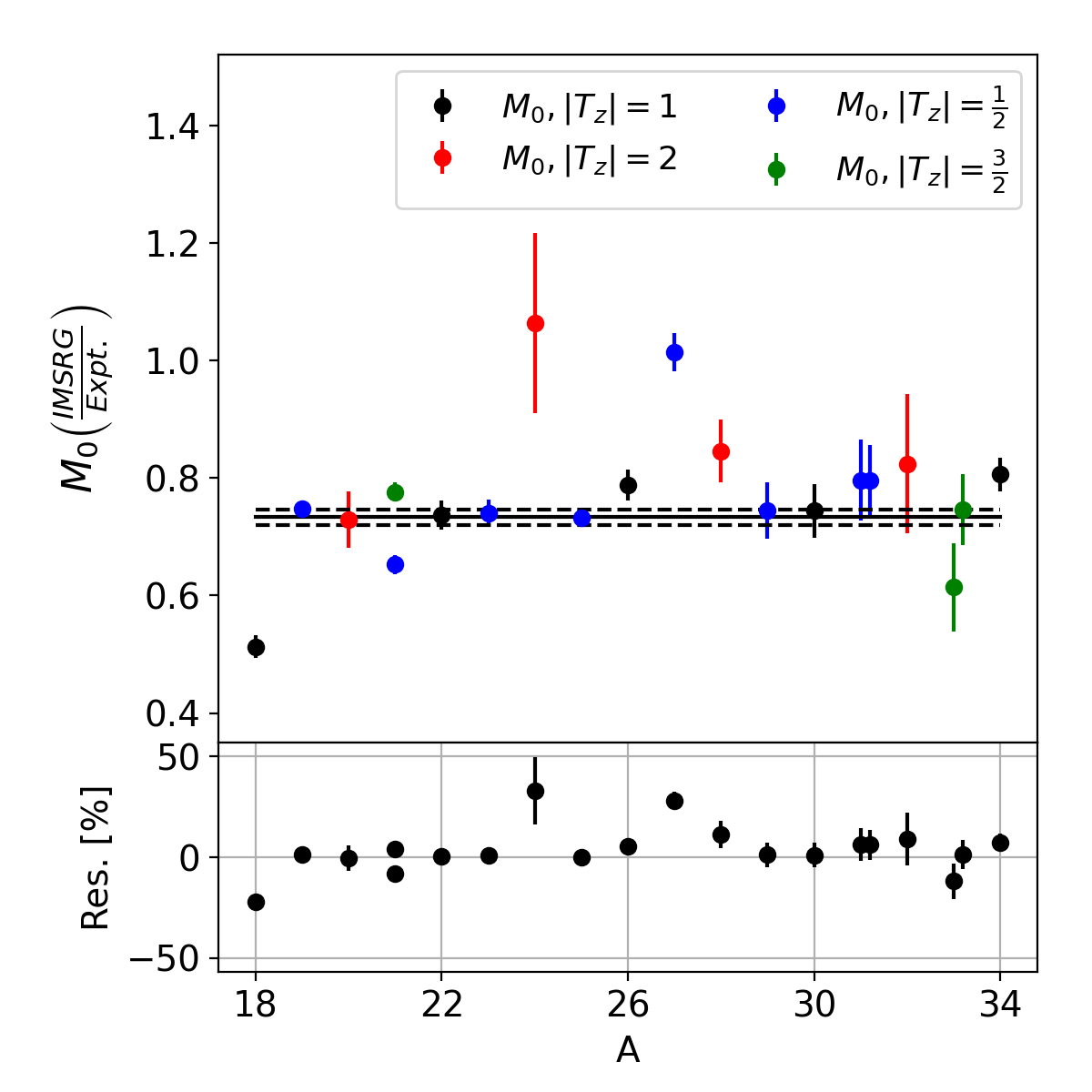}}
\caption{(Top) Ratio of $M_0$ calculated from VS-IMSRG calculations and extracted from experimental data for $sd$-shell nuclei. A fit to a constant (solid line) is also shown, along with the $1\sigma$ uncertainties (dashed lines) on the result. The fit yields a deficiency in the VS-IMSRG $M_0$ value of approximately $75\%$. (Bottom) Residuals for the fit.}
\label{fig:M0_IMSRG}
\end{figure}

\begin{figure}
\centerline{\includegraphics[width=\linewidth]{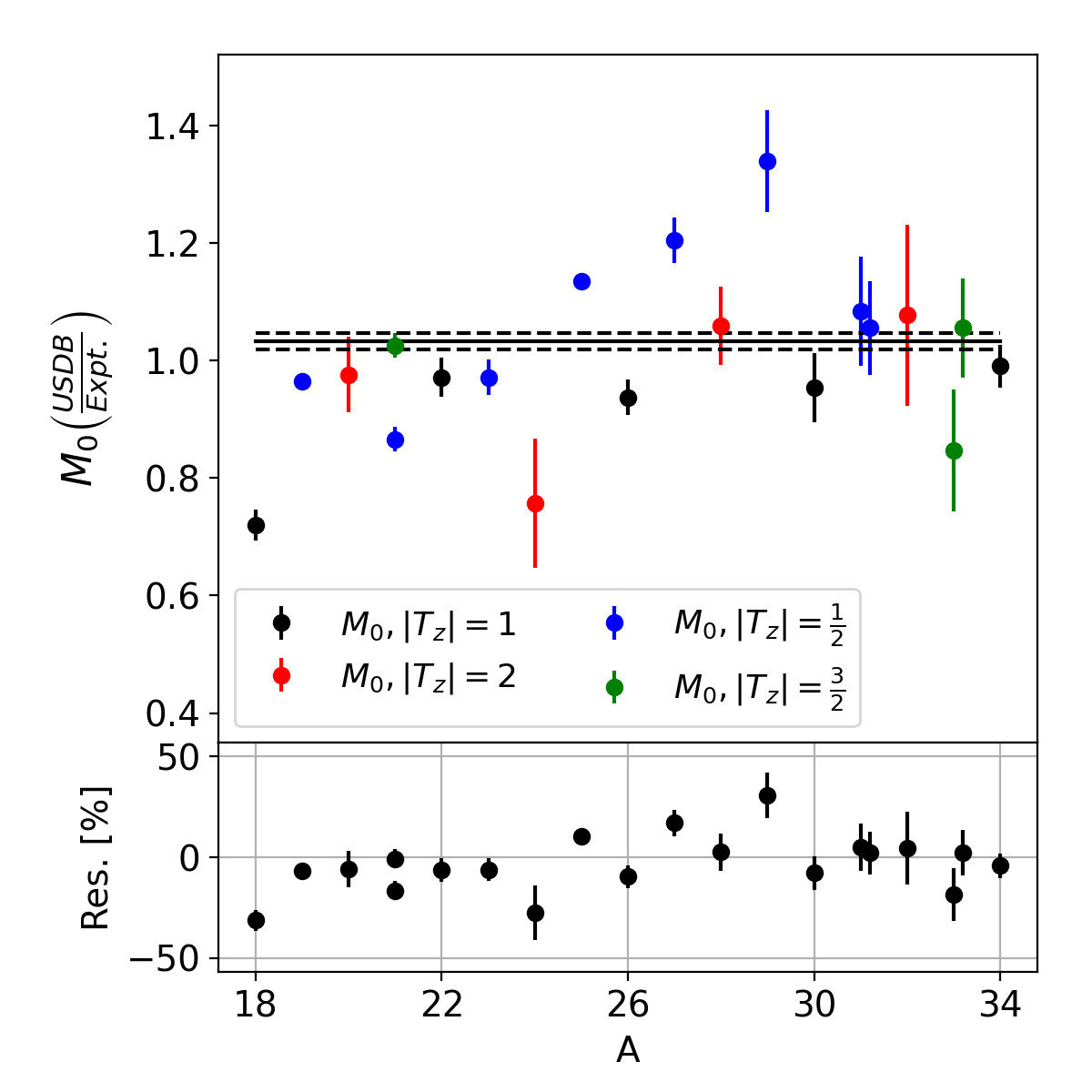}}
\caption{(Top) Ratio of $M_0$ calculated from USDB calculations and extracted from experimental data for $sd$-shell nuclei. A fit to a constant (solid line) is also shown, along with the $1\sigma$ uncertainties (dashed lines) on the result. The fit indicates that the USDB $M_0$ values approximately consistent with experiment on average, with $\frac{M_0 (USDB)}{M_0 Expt.} = 1.032 (26)$. (Bottom) Residuals for the fit. }
\label{fig:M0_USDB}
\end{figure}

To investigate this effect, we employ the isoscalar ($M_0$) and isovector ($M_1$) matrix element formalism of Brown et al.~\cite{ref:Brown_82}, where
\begin{equation}
	M_0 = \frac{\sqrt{B(E2;T_z < 0)} + \sqrt{B(E2;T_z > 0)}}{2},
\end{equation}
and
\begin{equation}
	M_1 = \left|\frac{\sqrt{B(E2;T_z < 0)} - \sqrt{B(E2;T_z > 0)}}{\Delta T_z}\right|.
	\label{eq:M1}
\end{equation}
Ratios of experimental to theoretical $M_0$ values are shown in Fig.~\ref{fig:M0_IMSRG} for the VS-IMSRG calculations and in Fig.~\ref{fig:M0_USDB} for the USDB results. As expected, the VS-IMSRG results underpredict the experimental data, with $\frac{M_0 (IMSRG)}{M_0 (Expt.)} \approx 75\%$ on average. By comparison, on average the USDB calculations reproduce $M_0$ well, with a modest overprediction. Of note is that, while the VS-IMSRG calculations are unable to reproduce the $M_0$ values, they provide a slightly improved description to the shell-model calculations, as highlighted by the reduced scatter in the residuals shown in the bottom panels of Fig.~\ref{fig:M0_IMSRG} and Fig.~\ref{fig:M0_USDB}. Note, for comparison, that without the use of effective charges the USDB calculations yield $\frac{M_0 (USDB)}{M_0 (Expt.)} \approx 55\%$. 

\begin{figure}
\centerline{\includegraphics[width=\linewidth]{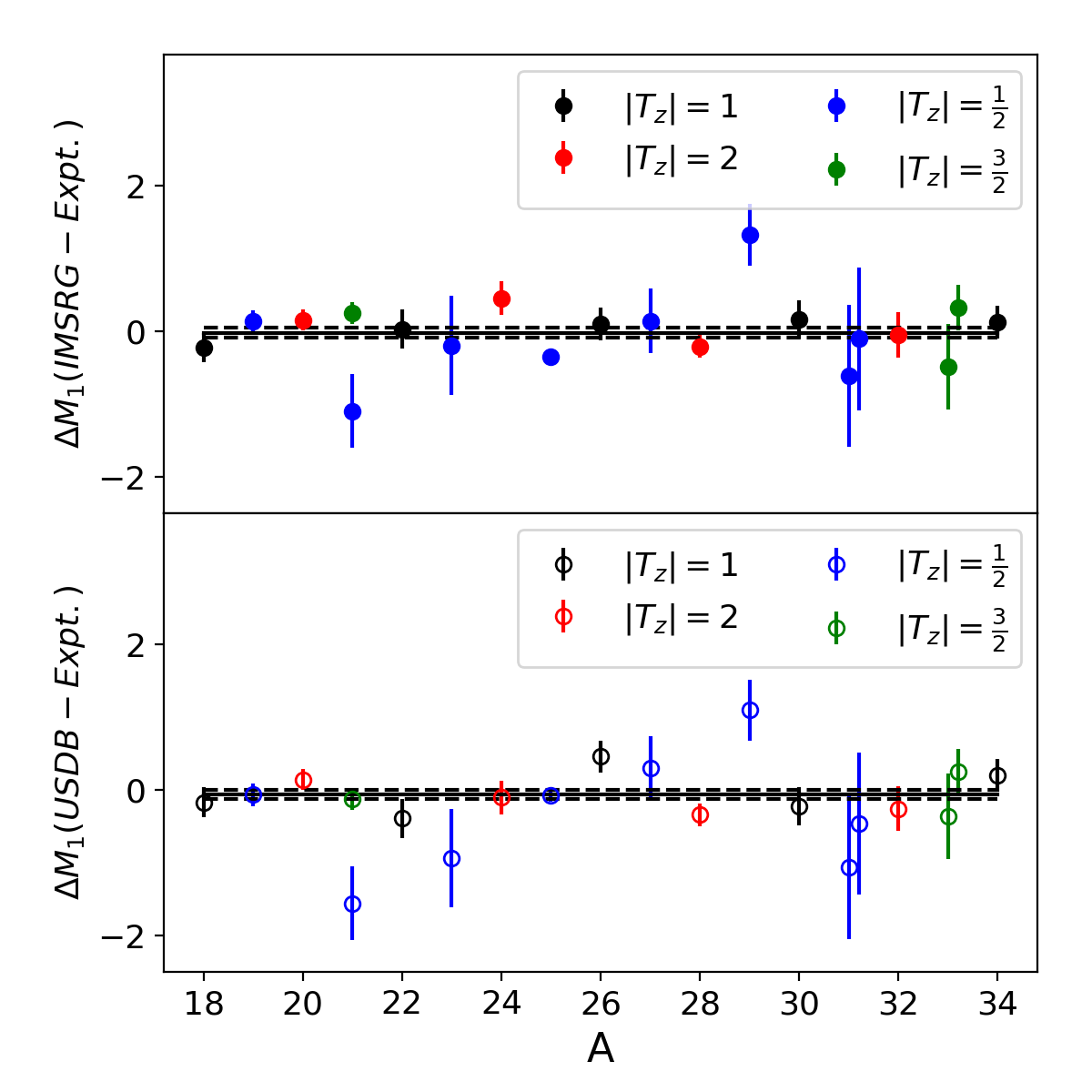}}
\caption{Differences between experimentally determined $M_1$ values and those calculated using the VS-IMSRG (top) and shell-model with USDB (bottom). A fit to a constant (solid) line is also shown, along with the $1\sigma$ uncertainties. Both USDB and VS-IMSRG results are consistent with zero - indicating no missing isovector contribution at the level of the presently achieved uncertainties.}
\label{fig:M1}
\end{figure}

Figure~\ref{fig:M1} shows the differences between experimental and calculated $|M_1|$ values for both the VS-IMSRG and USDB calculations. Both VS-IMSRG and USDB calculations yield, on average, $|M_1|$ values consistent with experiment. There are some notable deviations, in particular in the $A=29$, $\left|T_z\right|=\frac{1}{2}$ case. Here, the experimental data are rooted in a single excited-state lifetime measurement in \textsuperscript{29}Si~\cite{ref:Scherpenzeel_80}, and a pair of mixing ratio measurements~\cite{ref:Okano_60,ref:Byrski_74} for the decay of the first-excited state in \textsuperscript{29}P. Based on the significant deviation in isovector matrix elements found here for both USDB and VS-IMSRG calculations, these decays may merit further experimental study. This indicates that, to the level of presently available experimental and theoretical uncertainty, the missing strength in the VS-IMSRG calculations is isoscalar, in line with na\"ive expectations. In previous work~\cite{ref:Henderson_18} we noted a potential isovector component to the missing VS-IMSRG strength from a study of $|T_z|=1$ nuclei. In light of the present, comprehensive study of the $sd$-shell, however, we find that this effect is in fact due to a deficiency of isoscalar strength, rather than any excess of isovector. 

One can estimate the contribution of the Hamiltonian to the deficiency in isoscalar strength through comparison with charge-radii. As previously mentioned~\cite{ref:Simonis_17} it has been found that VS-IMSRG calculations using the EM1.8/2.0 interaction underpredict charge-radii, with the underprediction being approximately 7\%. The corresponding underprediction of the $E2$ matrix element would then be 13\%. Clearly, as shown in Fig.~\ref{fig:M0_IMSRG}, the underprediction in the calculations is larger than can be explained by the small radii. 


Prior to further discussion of the origins of the missing $E2$ strength it is worth recalling the text book picture, exemplified by the book of Bohr and Mottelson~\cite{ref:BohrMottelson}, in which effective charges arise primarily due to coupling to a giant resonance. In that work, the $E2$ effective charge is estimated with particle-vibration coupling in first order perturbation theory.
The vibration is described in the random phase approximation (RPA), obtained with a schematic quadrupole-quadrupole interaction.
To the extent that this treatment captures the dominant collective enhancement, any model that incorporates these degrees of freedom should successfully reproduce $E2$ strength. The present IMSRG method, as well as the coupled-cluster method introduced shortly, naturally capture all RPA correlations. We also comfortably include the most important $\Delta N=2$ excitations, thanks to the 13 major oscillator shells employed in the IMSRG evolution.

Inspired by the treatment in Ref.~\cite{ref:BohrMottelson}, we estimate the effective charge for a $d_{5/2}$ nucleon above $^{16}$O using a schematic model, consisting of a harmonic oscillator single-particle spectrum and a quadrupole-quadrupole residual interaction (see Appendix~\ref{app:SchematicModel} for more detail), working in a space $e_{\rm max}=3$, i.e. all orbits up to and including the $pf$ shell.
We compute the effective charge  at first-order in the particle-vibration coupling.
We treat the vibration at three levels of approximation: ``core polarization", corresponding to non-interacting particle-hole excitations; Tamm-Dancoff approximation (TDA); and the RPA.
See Appendix~\ref{app:RPA} for more details.
We also compute the effective charge with the VS-IMSRG by decoupling the $sd$ shell valence space.
Finally, we employ a direct configuration interaction (CI) diagonalization, performed with the KSHELL code~\cite{Shimizu2019} with a truncation on the number $N_{\rm max}$ of excitation quanta out of the naive ground state.
(We stop at $N_{\rm max}=6$ because the dimension for $N_{\rm max}=8$ is over $10^9$).
The results are listed in the top two rows of Table~\ref{tab:effcharge}.

\begin{table}[h]
    \caption{Effective charge for a $d_{5/2}$ nucleon above an $^{16}$O core for different approximation schemes and interaction models. The model space is defined by $e_{\rm max}=3$, $\hbar\omega=16$~MeV. The last three columns give results of direct diagonalization truncated to $N_{\rm max}$ quanta of excitation.}
    \label{tab:effcharge}
    \centering
    \begin{ruledtabular}
    \begin{tabular}{cr|cccc|ccc}
     & & & & & VS-  & \multicolumn{3}{c}{CI $N_{\rm max}$} \\
    int. &  & CP & TDA & RPA & IMSRG &  2 & 4 & 6 \\
    \hline
    $Q\cdot Q$&  $e_n$ & 0.23 & 0.29 & 0.42 & 0.43  & 0.26 & 0.32 & 0.41 \\ 
     & $e_p$ & 1.25 & 1.31 & 1.44 & 1.49  & 1.30 & 1.37 & 1.46 \\ 
    \hline
    NN  &  $e_n$ & 0.16 & 0.17 & 0.17 & 0.17 & 0.14 & 0.17 & 0.19 \\ 
         only   &  $e_p$ & 1.05 & 1.09 & 1.10 & 1.04  & 1.04 & 1.05 & 1.05 \\ 
   \hline
    NN & $e_n$& 0.24 & 0.31 & 0.33 & 0.26 &  0.20 & 0.23 & 0.29 \\ 
       +3N   & $e_p$& 1.07 & 1.16 & 1.19 & 1.02 & 1.04 & 1.05 & 1.05 \\ 
    \end{tabular}
    \end{ruledtabular}
\end{table}

We see that the effective charge obtained with the IMSRG is comparable to---in fact slightly larger than---that obtained with the RPA.
(We also note that the CI calculation does not show clear evidence of convergence in the $N_{\rm max}$ truncation).
It is clear that if the textbook picture were the full story, then the IMSRG should not dramatically underpredict $E2$ strength.
A natural suspicion is that the schematic interaction is not sufficiently realistic.

We repeat the exercise, replacing the schematic Hamiltonian with the chiral NN interaction plus the intrinsic kinetic energy in the $e_{\rm max}=3$, $\hbar\omega=16$~MeV space.
We also include the chiral 3N interaction, normal ordered with respect to the Hartree-Fock $^{16}$O reference obtained within that model space.
(Because we work in the Hartree-Fock basis, the $N_{\rm max}$ truncation does not strictly count oscillator quanta).
The results are given in Table~\ref{tab:effcharge}.
Here, we observe that the IMSRG yields a \emph{smaller} effective charge than the RPA, especially for protons.
This indicates that other non-RPA contributions act to reduce the effective charge,
consistent with the findings of Siegel and Zamick~\cite{ref:Siegel_70} at second order in perturbation theory.

In fact, only a subset of the terms in the IMSRG(2) flow equation contribute to RPA topologies.
These are the $[\eta_{\rm 2b},\mathcal{O}_{\rm 1b}]_{\rm 1b}$ part\footnote{The subscripts 1b and 2b denote the one-body and two-body parts, respectively, of the given operator.} of the flow equation for $\mathcal{O}(s)$, and the particle-hole part of the $[\eta_{\rm 2b},H_{\rm 2b}]_{\rm 2b}$ flow equation for $H(s)$ which in turn defines $\eta(s)$.
If we neglect all terms other than those mentioned above, the IMSRG effective charges for the NN+3N interaction increase to $e_n=0.28$ and $e_p =1.14$.
On the other hand, for the schematic $Q\cdot Q$ interaction, the effective charges are only slightly modified to $e_n=0.43$ and $e_p=1.40$.

The RPA effective charges includes all TDA contributions as well as 2p2h, 4p4h, 6p6h, etc. excitations in the ``zero-phonon'' $^{16}$O ground state, leading to 1p1h, 3p3h, 5p5h, etc. excitations in the ``one-phonon'' excited states.
The difference between TDA and RPA is an indication of the importance of the RPA ground-state correlations.
While these correlations have a large effect with the schematic $Q\cdot Q$ interaction, they appear to be much less important with the realistic NN+3N interaction.

The $N_{\max}$=2 CI effective charge includes all TDA contributions, as well as 2p2h excitations from the $p$ shell to the $sd$ shell and terms where both the initial and final state have 1p1h or 2p2h excitations.
In addition, the CI wave function is normalized, while the TDA wave function is only normalized to first order in the particle-vibration coupling.
The difference between these two columns indicates the importance of these additional effects, which are also missing in the RPA.
For the schematic and NN-only interactions, these corrections lead to a minor reduction, while for the NN+3N interaction the effect is more dramatic.


More insight may be gleaned by inspecting the individual contribution of each operator matrix element $\mathcal{O}_{ab} $ to the effective charge for each combination of orbits $a$ and $b$.
This is summarized in Table~\ref{tab:OBTD} for each oscillator shell sub-block (e.g. the $p\to pf$ row is the sum of all contributions $\mathcal{O}_{ab}+\mathcal{O}_{ba}$ with $a$ in the $0p$ shell and $b$ in the $1p0f$ shell).
By construction, the RPA effective charges only receive contributions from $s\to sd$ and $p\to pf$.
The IMSRG and CI calculations, however, also receive contributions from the ``diagonal'' $p\to p$, $sd\to sd$, $pf\to pf$ terms.
Because the comparison with experiment indicates that missing IMSRG strength is of isoscalar nature, we also present in Table~\ref{tab:OBTD} the contributions to the isoscalar and isovector effective charges.

\begin{table}[]
    \caption{Contribution to the $0d_{5/2}$ effective charges broken down by sub-block of the operator $\mathcal{O}_{ab}$. We use the EM1.8/2.0 NN+3N interaction at $\hbar\omega=16$, $e_{\rm max}=3$. The CI calculations correspond to $N_{\rm max}=6$. The last row gives the summed isovector and isoscalar contributions.}
    \label{tab:OBTD}
\begin{ruledtabular}
    \centering
    \begin{tabular}{c|rrr|rrr}
     & \multicolumn{3}{c|}{$\delta e_n$} & \multicolumn{3}{c}{$\delta e_p$} \\
         & RPA & IMSRG & CI  &RPA & IMSRG & CI  \\
         \hline
$p\to pf$  &  0.248  & 0.192 & 0.175  &  0.135  & 0.051  & 0.058 \\ 
$s\to sd$  &  0.085  & 0.071 & 0.075  &  0.051  & 0.031  & 0.033 \\ 
$p\to p$   &  0      & 0.002 & 0.023  &  0      & 0.011  & 0.023 \\ 
$sd\to sd$ &  0      & -0.014 & 0.001 &  0      & -0.083 & -0.074 \\ 
$fp\to fp$ &  0      & 0.012 & 0.012  &  0      & 0.009  & 0.009  \\
\hline 
     & \multicolumn{3}{c|}{$\delta e_n+\delta e_p$} & \multicolumn{3}{c}{$\delta e_n-\delta e_p$} \\
     \hline
     $p\to pf$  &  0.383  & 0.243 & 0.233  &  0.112  & 0.141  & 0.116 \\ 
$s\to sd$  &  0.137  & 0.102 & 0.108  &  0.034  & 0.040  & 0.042 \\ 
$p\to p$   &  0      & 0.014 & 0.046  &  0      & -0.009  & 0.000 \\ 
$sd\to sd$ &  0      & -0.097 & -0.073 &  0      & 0.070 & 0.075 \\ 
$fp\to fp$ &  0      & 0.021 & 0.021  &  0      & 0.003  & 0.003  \\
$\sum$ & 0.520 & 0.283 & 0.335 &  0.146 & 0.245 & 0.235
    \end{tabular}
\end{ruledtabular}
\end{table}

We see that, compared to the CI result, the RPA overestimates the isoscalar effective charge and underestimates the isovector charge.
The IMSRG underestimates the isoscalar charge and slightly overestimates the isovector charge, with the largest discrepancy coming from the isoscalar $p\to p$ sub-block.

Because the $p$ orbits are occupied in the zero-order wave function, a $p\to p$ transition requires particle-hole excitations in both the initial and final state.
In the particle-vibration coupling picture, this would correspond to terms which are at least second order in the particle-vibration interaction.
If we restrict the CI calculation to $N_{\rm max}=4$, the isoscalar $p\to p$ contribution is reduced by nearly half to 0.027, indicating the importance of either 3p3h $p\to pf$ excitations, or 6p6h $p\to sd$ (or something inbetween).
Repeating the calculation with the $fp$ shell excluded leaves the $p\to p$ contribution essentially unchanged.
Finally, with the $pf$ shell excluded we are able to extend the CI calculation to $N_{\rm max}=8$, where we find that the $p\to p$ contribution is essentially unchanged from the $N_{\rm max}=6$ calculation.
We conclude that the $p\to p$ term has a significant contribution from 6p6h $p\to sd$ excitations, while 8p8h are less important.

As noted above, our CI calculations are not fully converged with respect to the $N_{\rm max}$ truncation, and so it is not clear how much the IMSRG result deviates from the exact one.
To address this, we consider a more tractable problem where an exact diagonalization is possible, namely \textsuperscript{14}C, using the \textit{psdmwk} interaction~\cite{ref:Warburton_92PRC,ref:Warburton_92}.


The choice of \textsuperscript{14}C also enables a comparison with coupled-cluster (CC).
Full configuration interaction (FCI) calculations were performed in NuShellX. Quadrupole transition amplitudes were therefore determined for protons and neutrons ($A_p$ and $A_n$, respectively) such that
\begin{equation}
B(E2;2^+\rightarrow0^+) = \frac{(A_p e_p + A_n e_n)^2}{5},
\end{equation}
where $e_p$ and $e_n$ correspond to the proton and neutron effective charges, respectively.
Because we use a Hamiltonian with phenomonologically determined matrix elements, the corresponding radial wave functions are arbitrary.
We use a harmonic oscillator basis and present the amplitudes in units of the oscillator length squared, $b^2$.
In the VS-IMSRG calculations, the IMSRG transformation is used to decouple the $p$-shell from the $sd$-shell, and the resulting $p$-shell interaction is diagonalized. We also present CC calculations in which the \textsuperscript{14}C $2^+$ excited state is computed using the equation-of-motion coupled-cluster (EOM-CC) formalism~\cite{ref:Stanton_93} which amounts to an expansion in particle-hole excitations out of the CC solution for the $0^+$ ground state. The order of the expansions used is denoted in parentheses with the first value indicating the highest order ground-state expansion and the second indicating the EOM expansion used to calculate the excited state. Ground-state expansions are CCSD, CCSDT-1 and CCSDT-3, corresponding to singles-doubles, singles-doubles and leading-order triples, and singles-doubles and up to third-order triples~\cite{ref:Lee_84,ref:Watts_96}, respectively. The order of the excited-state expansion is given as S, D or T for 1p-1h, 2p-2h and 3p-3h expansions out of the ground-state, respectively. For example, CC(D/S) corresponds to a CCSD ground-state with the $2^+$ state expanded in terms of 1p-1h excitations, and CC(T-1/T) corresponds to a CCSDT-1 ground-state with the $2^+$ state expanded in terms of excitations up to 3p-3h.

\begin{figure}
\centerline{\includegraphics[width=\linewidth]{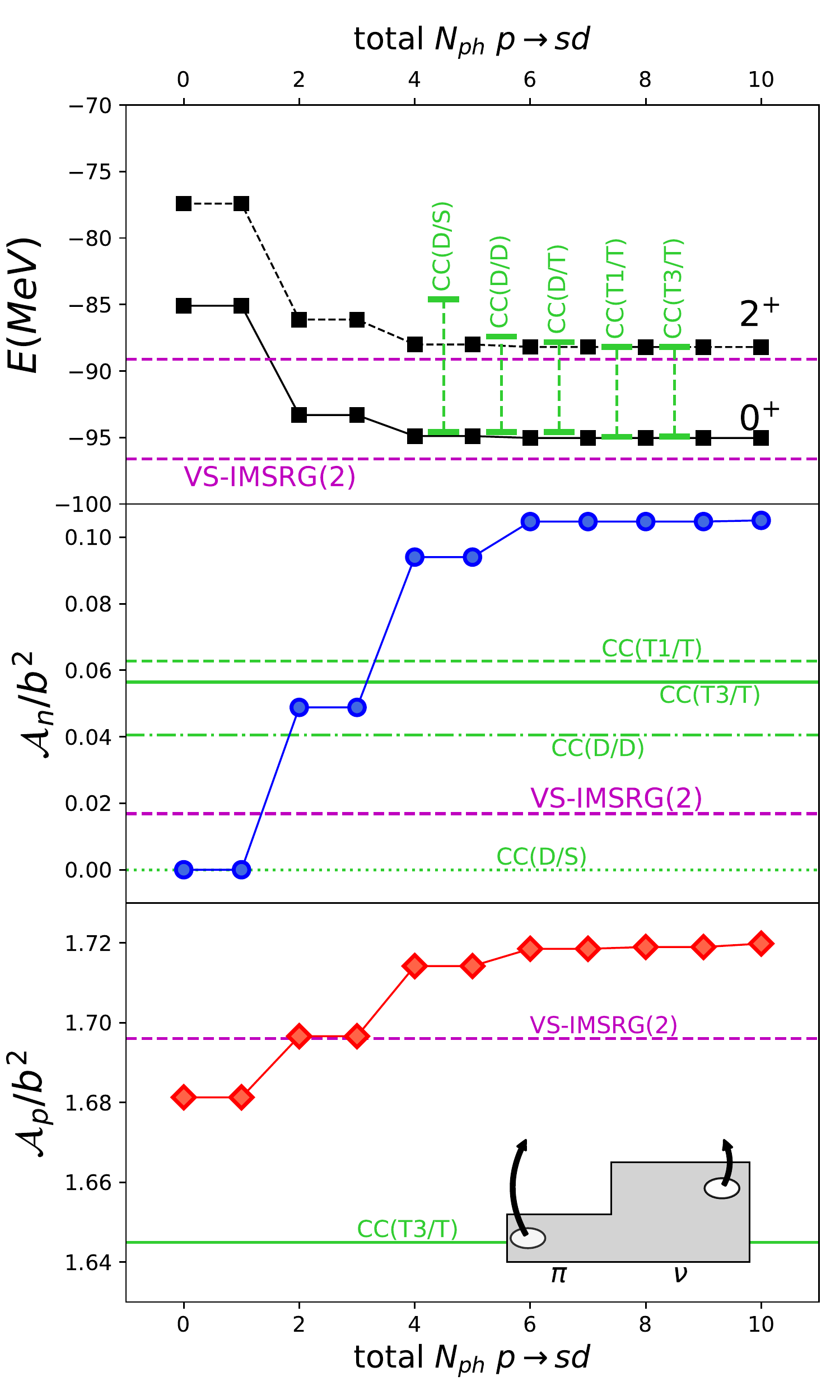}}
\caption{Absolute $0^+_1$ and $2^+_1$ state energies (top row), neutron quadrupole-excitation amplitudes (middle row) and proton quadrupole-excitation amplitudes (bottom row) for $^{14}$C calculated in a full configuration interaction diagonalization with the \textit{psdmwk} interaction, plotted against the number of excitations permitted from the $p$ to $sd$ shell. Also shown are the values calculated using the VS-IMSRG method, as well as using an equation-of-motion coupled-cluster methodology. See text for details of calculations.}
\label{fig:phexcitations}
\end{figure}

It is found that the FCI ground- and excited state energies are already well reproduced at the EOM-CCSD level and by the VS-IMSRG calculations. On the other hand, we find that the quadrupole amplitudes are consistently underpredicted by VS-IMSRG, consistent with the comparison with experimental data. The EOM-CC calculations show improved reproduction of the FCI interactions with increasing order of expansion, but still significantly underpredict the quadrupole amplitudes even at the CC(T-3/T) order. Since all calculations were performed with the same initial Hamiltonian, this missing strength must arise from the many-body approximation. To further investigate this underprediction, the microscopic behaviour of the configuration interaction was controlled using a series of CI calculations, with a truncation on the total number of nucleon excitations out of the $p$ shell. Figure~\ref{fig:phexcitations} shows the results of these calculations. 

The VS-IMSRG calculations yield a larger quadrupole amplitude than that obtained with 0p0h (i.e. $p$-shell only) calculations. This is as expected: the VS-IMSRG evolution approximately decouples the $p$ shell from the $sd$ shell and then diagonalizes within the $p$ shell, so the amplitude within the $p$ shell should be completely accounted for. As the truncation conditions are relaxed and excitations into the $sd$ shell are permitted, the VS-IMSRG calculations soon fail to capture the additional strength, indicating that the SRG-decoupling of the spaces has resulted in information relevant to the quadrupole amplitude being lost. The CC calculations do not fully reproduce the pure p-shell amplitude for the open-shell protons (bottom panel, Fig.~\ref{fig:phexcitations}). Perhaps more interestingly, however, is a comparison to the closed-shell neutrons (middle panel, Fig.~\ref{fig:phexcitations}). Increasing the order of the CC-EOM calculations - effectively allowing for correlated 3p-3h excitations, in the case of CC(T/T) - clearly helps restore missing strength. Due to the exponential ansatz, the CC wavefunctions at CC(T/T) contain a fraction of up to 10p-10h excitations, however beyond 3p-3h these excitations are limited to disconnected cluster terms, i.e. products of lower rank excitations.
The above, combined with the observation that most of the strength is reproduced at the 4p-4h level, suggests that accurate treatment of $E2$ observables in the IMSRG or CC frameworks will require including at least four-body operators.

While a straightforward inclusion of four-body operators would lead to an intractable $n^{12}$ scaling, where $n$ is the number of single-particle states, one would hope that the relevant contributions could be captured within a manageable approximation scheme.
For example, one could imagine including effective phonon degrees of freedom.
The present work also suggests that additional isovector degrees of freedom would be less important.

\section{Conclusions}

Consistent with previous work, it was found that the VS-IMSRG calculations significantly underpredict the $E2$ transition strength. This underprediction must arise from the two-body truncation to the operator evolution applied to make the method computationally tractable. A comprehensive survey of the literature was performed and compared with the VS-IMSRG and shell-model results. It is found that the missing strength is predominantly isoscalar in nature, while the missing isovector contribution is consistent with zero within presently available experimental and theoretical uncertainties. We provided a brief discussion on the VS-IMSRG calculations, presenting a potential explanation for the apparent lack of isovector contribution to the missing $E2$ strength.

Inspired by textbook descriptions of effective charges, we compared VS-IMSRG calculations with a number of approximations, including the RPA, using both a schematic quadrupole-quadrupole interaction and the chiral interactions used previously. We found that the effective charge calculated from the VS-IMSRG exceeded those of the approximate methods for the schematic interaction. In the case of the chiral interactions, we found evidence of non-RPA contributions acting to reduce the effective charge for the VS-IMSRG, while also showing that the effect of including 3N forces when employing the RPA was effectively negligible.
Finally, full- and truncated-configuration interaction calculations of \textsuperscript{14}C were performed and compared with both VS-IMSRG and coupled-cluster calculations.
These calculations help identify that connected 4p-4h excitations, missing in the VS-IMSRG and CC calculations and typically unimportant for energies, need to be included for accurate reproduction of quadrupole observables.

\section{Acknowledgements}

This work has been supported by the Natural Sciences and Engineering Research Council of Canada (NSERC), The Canada Foundation for Innovation and the British Columbia Knowledge Development Fund. TRIUMF receives federal funding via a contribution agreement through the National Research Council of Canada. Computations were performed with an allocation of computing resources on Cedar at WestGrid and Compute Canada, and on the Oak Cluster at TRIUMF managed by the University of British Columbia department of Advanced Research Computing (ARC). Work at LLNL was performed under contract DE-AC52-07NA27344. This work was supported by the Office of Nuclear Physics, U.S. Department of Energy, under grants desc0018223 (NUCLEI SciDAC-4 collaboration) and by the Field Work Proposal ERKBP72 at Oak Ridge National Laboratory (ORNL).
SRS was supported by the U.S. Department of Energy office of Science, Office of Nuclear Physics, under contract DE-FG02-97ER41014 and DEAC02-06CH11357. JH is supported at the University of Surrey under UKRI Future Leaders Fellowship grant no. MR/T022264/1.

The codes imsrg++~\cite{ref:imsrgcode} and nutbar~\cite{ref:nutbar} used in this work make use of the Armadillo library~\cite{ref:Armadillo}.

\appendix

\section{Schematic model for effective charge calculation\label{app:SchematicModel}}
We take a harmonic oscillator single-particle spectrum with a residual quadrupole-quadrupole interaction.
Specifically, the Hamiltonian is
\begin{equation}
    H = \sum_i \epsilon_i a^{\dagger}_ia_i + \frac{1}{4}\sum_{ijkl} V_{ijkl} a^{\dagger}_ia^{\dagger}_ja_la_k
\end{equation}
where $\epsilon_i=2n_i+\ell_i$ and 
\begin{equation}
    V_{ijkl} = \kappa\left( Q_{ik}Q_{jl} - Q_{il}Q_{jk}\right).
\end{equation}
Here $Q$ is the quadrupole operator $r^2 Y_2(\theta,\phi)$, and $\kappa$ is a strength parameter.
Following the estimate in \cite{ref:BohrMottelson}, we take
\begin{equation}
\kappa = -\xi \frac{4\pi}{5} \frac{M\omega^2}{A\langle r^2\rangle},
\end{equation}
with $\xi=1$, well below the critical value $\xi_{\rm crit}\approx 9/4$ at which the RPA solution for excitation energy collapses, signaling instability against static deformation.

\section{RPA effective charge calculations\label{app:RPA}}

In this section we outline the calculation of effective charges in the random phase approximation (see e.g.~\cite{ref:Siegel_70}).
We use the convention that $a,b,c$ label particles, $i,j,k$ label holes, and $p,q$ run over all orbits.
In a $J$-coupled representation, the RPA effective operator corresponding to  
a one-body operator $\mathcal{O}^{\lambda}$ of spherical-tensor rank $\lambda$ with reduced matrix elements $O^{\lambda}_{pq}=\langle p \| \mathcal{O}^{\lambda} \| q\rangle$ is given by
~\cite{ref:Ring_80}
\begin{equation}\label{eq:CorePol}
     O^{\lambda,{\rm eff}}_{ab} =  O^{\lambda}_{ab} + \sum_{pq} (n_q-n_p)   \frac{\bar{V}^{\lambda}_{a\bar{b}p\bar{q}}}{\Delta_{qapb}} O^{\lambda,{\rm eff}}_{pq}
    .
\end{equation}
where we have used a shorthand for the energy denominator $\Delta_{pars} \equiv \epsilon_p+\epsilon_q-\epsilon_r-\epsilon_s$, $n_p$ is the occupation of orbit $p$,
and the Pandya-transformed interaction matrix elements
\begin{equation}
    \bar{V}^{\lambda}_{p\bar{q}r\bar{s}}
    = -\sum_J (2J+1) \begin{Bmatrix} j_p & j_q & \lambda \\ j_r & j_s & J\end{Bmatrix} V^{J}_{psrq}.
\end{equation}
The bar on an index denotes a time-reversed state.
Next, we form the matrix $\mathbb{M}_{ab}$
    with elements
    \begin{equation}\label{eq:Mmatrix}
    \langle p\bar{q} \lambda | \mathbb{M}_{ab} | r\bar{s}\lambda\rangle = (n_s-n_r) \frac{\bar{V}^{\lambda}_{p\bar{q}r\bar{s}}}{\Delta_{sarb}}
    \end{equation}
and the vector $\vec{O}$ of matrix elements, in terms of which \eqref{eq:CorePol} becomes
\begin{equation}
    O^{\lambda,{\rm eff}}_{ab} = \left[\vec{O} + \mathbb{M}_{ab}\vec{O}^{\rm eff}\right]_{ab}.
\end{equation}
The RPA resummation is performed by matrix inversion
\begin{equation}
    O^{\lambda,{\rm eff}}_{ab}
     =  \left[ \left( \mathbb{I} - \mathbb{M}_{ab}\right)^{-1}
     \vec{O} \right]_{ab}
\end{equation}
TDA corresponds to \eqref{eq:Mmatrix} multiplied by $\delta_{n_p n_r}\delta_{n_qn_s}$, and first-order core-polarization is obtained from \eqref{eq:CorePol} with the replacement $O^{\lambda,{\rm eff}}\to O^{\lambda}$ on the right hand side.
\bibliography{mg23}

\begin{thebibliography}{43}%
\makeatletter
\providecommand \@ifxundefined [1]{%
 \@ifx{#1\undefined}
}%
\providecommand \@ifnum [1]{%
 \ifnum #1\expandafter \@firstoftwo
 \else \expandafter \@secondoftwo
 \fi
}%
\providecommand \@ifx [1]{%
 \ifx #1\expandafter \@firstoftwo
 \else \expandafter \@secondoftwo
 \fi
}%
\providecommand \natexlab [1]{#1}%
\providecommand \enquote  [1]{``#1''}%
\providecommand \bibnamefont  [1]{#1}%
\providecommand \bibfnamefont [1]{#1}%
\providecommand \citenamefont [1]{#1}%
\providecommand \href@noop [0]{\@secondoftwo}%
\providecommand \href [0]{\begingroup \@sanitize@url \@href}%
\providecommand \@href[1]{\@@startlink{#1}\@@href}%
\providecommand \@@href[1]{\endgroup#1\@@endlink}%
\providecommand \@sanitize@url [0]{\catcode `\\12\catcode `\$12\catcode
  `\&12\catcode `\#12\catcode `\^12\catcode `\_12\catcode `\%12\relax}%
\providecommand \@@startlink[1]{}%
\providecommand \@@endlink[0]{}%
\providecommand \url  [0]{\begingroup\@sanitize@url \@url }%
\providecommand \@url [1]{\endgroup\@href {#1}{\urlprefix }}%
\providecommand \urlprefix  [0]{URL }%
\providecommand \Eprint [0]{\href }%
\providecommand \doibase [0]{http://dx.doi.org/}%
\providecommand \selectlanguage [0]{\@gobble}%
\providecommand \bibinfo  [0]{\@secondoftwo}%
\providecommand \bibfield  [0]{\@secondoftwo}%
\providecommand \translation [1]{[#1]}%
\providecommand \BibitemOpen [0]{}%
\providecommand \bibitemStop [0]{}%
\providecommand \bibitemNoStop [0]{.\EOS\space}%
\providecommand \EOS [0]{\spacefactor3000\relax}%
\providecommand \BibitemShut  [1]{\csname bibitem#1\endcsname}%
\let\auto@bib@innerbib\@empty
\bibitem [{\citenamefont {Henderson}\ \emph {et~al.}(2018)\citenamefont
  {Henderson}, \citenamefont {Hackman}, \citenamefont {Ruotsalainen},
  \citenamefont {Stroberg}, \citenamefont {Launey}, \citenamefont {Holt},
  \citenamefont {Ali}, \citenamefont {Bernier}, \citenamefont {Bentley},
  \citenamefont {Bowry}, \citenamefont {Caballero-Folch}, \citenamefont
  {Evitts}, \citenamefont {Frederick}, \citenamefont {Garnsworthy},
  \citenamefont {Garrett}, \citenamefont {Jigmeddorj}, \citenamefont {Kilic},
  \citenamefont {Lassen}, \citenamefont {Measures}, \citenamefont {Muecher},
  \citenamefont {Olaizola}, \citenamefont {O'Sullivan}, \citenamefont
  {Paetkau}, \citenamefont {Park}, \citenamefont {Smallcombe}, \citenamefont
  {Svensson}, \citenamefont {Wadsworth},\ and\ \citenamefont
  {Wu}}]{ref:Henderson_18}%
  \BibitemOpen
  \bibfield  {author} {\bibinfo {author} {\bibfnamefont {J.}~\bibnamefont
  {Henderson}}, \bibinfo {author} {\bibfnamefont {G.}~\bibnamefont {Hackman}},
  \bibinfo {author} {\bibfnamefont {P.}~\bibnamefont {Ruotsalainen}}, \bibinfo
  {author} {\bibfnamefont {S.}~\bibnamefont {Stroberg}}, \bibinfo {author}
  {\bibfnamefont {K.}~\bibnamefont {Launey}}, \bibinfo {author} {\bibfnamefont
  {J.}~\bibnamefont {Holt}}, \bibinfo {author} {\bibfnamefont {F.}~\bibnamefont
  {Ali}}, \bibinfo {author} {\bibfnamefont {N.}~\bibnamefont {Bernier}},
  \bibinfo {author} {\bibfnamefont {M.}~\bibnamefont {Bentley}}, \bibinfo
  {author} {\bibfnamefont {M.}~\bibnamefont {Bowry}}, \bibinfo {author}
  {\bibfnamefont {R.}~\bibnamefont {Caballero-Folch}}, \bibinfo {author}
  {\bibfnamefont {L.}~\bibnamefont {Evitts}}, \bibinfo {author} {\bibfnamefont
  {R.}~\bibnamefont {Frederick}}, \bibinfo {author} {\bibfnamefont
  {A.}~\bibnamefont {Garnsworthy}}, \bibinfo {author} {\bibfnamefont
  {P.}~\bibnamefont {Garrett}}, \bibinfo {author} {\bibfnamefont
  {B.}~\bibnamefont {Jigmeddorj}}, \bibinfo {author} {\bibfnamefont
  {A.}~\bibnamefont {Kilic}}, \bibinfo {author} {\bibfnamefont
  {J.}~\bibnamefont {Lassen}}, \bibinfo {author} {\bibfnamefont
  {J.}~\bibnamefont {Measures}}, \bibinfo {author} {\bibfnamefont
  {D.}~\bibnamefont {Muecher}}, \bibinfo {author} {\bibfnamefont
  {B.}~\bibnamefont {Olaizola}}, \bibinfo {author} {\bibfnamefont
  {E.}~\bibnamefont {O'Sullivan}}, \bibinfo {author} {\bibfnamefont
  {O.}~\bibnamefont {Paetkau}}, \bibinfo {author} {\bibfnamefont
  {J.}~\bibnamefont {Park}}, \bibinfo {author} {\bibfnamefont {J.}~\bibnamefont
  {Smallcombe}}, \bibinfo {author} {\bibfnamefont {C.}~\bibnamefont
  {Svensson}}, \bibinfo {author} {\bibfnamefont {R.}~\bibnamefont {Wadsworth}},
  \ and\ \bibinfo {author} {\bibfnamefont {C.}~\bibnamefont {Wu}},\ }\href@noop
  {} {\bibfield  {journal} {\bibinfo  {journal} {Physics Letters B}\ }\textbf
  {\bibinfo {volume} {782}},\ \bibinfo {pages} {468 } (\bibinfo {year}
  {2018})}\BibitemShut {NoStop}%
\bibitem [{\citenamefont {Tobin}\ \emph {et~al.}(2014)\citenamefont {Tobin},
  \citenamefont {Ferriss}, \citenamefont {Launey}, \citenamefont {Dytrych},
  \citenamefont {Draayer}, \citenamefont {Dreyfuss},\ and\ \citenamefont
  {Bahri}}]{ref:Tobin_14}%
  \BibitemOpen
  \bibfield  {author} {\bibinfo {author} {\bibfnamefont {G.~K.}\ \bibnamefont
  {Tobin}}, \bibinfo {author} {\bibfnamefont {M.~C.}\ \bibnamefont {Ferriss}},
  \bibinfo {author} {\bibfnamefont {K.~D.}\ \bibnamefont {Launey}}, \bibinfo
  {author} {\bibfnamefont {T.}~\bibnamefont {Dytrych}}, \bibinfo {author}
  {\bibfnamefont {J.~P.}\ \bibnamefont {Draayer}}, \bibinfo {author}
  {\bibfnamefont {A.~C.}\ \bibnamefont {Dreyfuss}}, \ and\ \bibinfo {author}
  {\bibfnamefont {C.}~\bibnamefont {Bahri}},\ }\href {\doibase
  10.1103/PhysRevC.89.034312} {\bibfield  {journal} {\bibinfo  {journal} {Phys.
  Rev. C}\ }\textbf {\bibinfo {volume} {89}},\ \bibinfo {pages} {034312}
  (\bibinfo {year} {2014})}\BibitemShut {NoStop}%
\bibitem [{\citenamefont {Navr\'atil}\ \emph {et~al.}(1997)\citenamefont
  {Navr\'atil}, \citenamefont {Thoresen},\ and\ \citenamefont
  {Barrett}}]{ref:Navratil_97}%
  \BibitemOpen
  \bibfield  {author} {\bibinfo {author} {\bibfnamefont {P.}~\bibnamefont
  {Navr\'atil}}, \bibinfo {author} {\bibfnamefont {M.}~\bibnamefont
  {Thoresen}}, \ and\ \bibinfo {author} {\bibfnamefont {B.~R.}\ \bibnamefont
  {Barrett}},\ }\href {\doibase 10.1103/PhysRevC.55.R573} {\bibfield  {journal}
  {\bibinfo  {journal} {Phys. Rev. C}\ }\textbf {\bibinfo {volume} {55}},\
  \bibinfo {pages} {R573} (\bibinfo {year} {1997})}\BibitemShut {NoStop}%
\bibitem [{\citenamefont {Stetcu}\ \emph {et~al.}(2005)\citenamefont {Stetcu},
  \citenamefont {Barrett}, \citenamefont {Navr\'atil},\ and\ \citenamefont
  {Vary}}]{ref:Stetcu_05}%
  \BibitemOpen
  \bibfield  {author} {\bibinfo {author} {\bibfnamefont {I.}~\bibnamefont
  {Stetcu}}, \bibinfo {author} {\bibfnamefont {B.~R.}\ \bibnamefont {Barrett}},
  \bibinfo {author} {\bibfnamefont {P.}~\bibnamefont {Navr\'atil}}, \ and\
  \bibinfo {author} {\bibfnamefont {J.~P.}\ \bibnamefont {Vary}},\ }\href
  {\doibase 10.1103/PhysRevC.71.044325} {\bibfield  {journal} {\bibinfo
  {journal} {Phys. Rev. C}\ }\textbf {\bibinfo {volume} {71}},\ \bibinfo
  {pages} {044325} (\bibinfo {year} {2005})}\BibitemShut {NoStop}%
\bibitem [{\citenamefont {Lisetskiy}\ \emph {et~al.}(2009)\citenamefont
  {Lisetskiy}, \citenamefont {Kruse}, \citenamefont {Barrett}, \citenamefont
  {Navratil}, \citenamefont {Stetcu},\ and\ \citenamefont
  {Vary}}]{ref:Lisetskiy_09}%
  \BibitemOpen
  \bibfield  {author} {\bibinfo {author} {\bibfnamefont {A.~F.}\ \bibnamefont
  {Lisetskiy}}, \bibinfo {author} {\bibfnamefont {M.~K.~G.}\ \bibnamefont
  {Kruse}}, \bibinfo {author} {\bibfnamefont {B.~R.}\ \bibnamefont {Barrett}},
  \bibinfo {author} {\bibfnamefont {P.}~\bibnamefont {Navratil}}, \bibinfo
  {author} {\bibfnamefont {I.}~\bibnamefont {Stetcu}}, \ and\ \bibinfo {author}
  {\bibfnamefont {J.~P.}\ \bibnamefont {Vary}},\ }\href {\doibase
  10.1103/PhysRevC.80.024315} {\bibfield  {journal} {\bibinfo  {journal} {Phys.
  Rev. C}\ }\textbf {\bibinfo {volume} {80}},\ \bibinfo {pages} {024315}
  (\bibinfo {year} {2009})}\BibitemShut {NoStop}%
\bibitem [{\citenamefont {Miyagi}\ \emph {et~al.}(2022)\citenamefont {Miyagi},
  \citenamefont {Stroberg}, \citenamefont {Navr\'atil}, \citenamefont
  {Hebeler},\ and\ \citenamefont {Holt}}]{ref:Miyagi_22}%
  \BibitemOpen
  \bibfield  {author} {\bibinfo {author} {\bibfnamefont {T.}~\bibnamefont
  {Miyagi}}, \bibinfo {author} {\bibfnamefont {S.~R.}\ \bibnamefont
  {Stroberg}}, \bibinfo {author} {\bibfnamefont {P.}~\bibnamefont
  {Navr\'atil}}, \bibinfo {author} {\bibfnamefont {K.}~\bibnamefont {Hebeler}},
  \ and\ \bibinfo {author} {\bibfnamefont {J.~D.}\ \bibnamefont {Holt}},\
  }\href {\doibase 10.1103/PhysRevC.105.014302} {\bibfield  {journal} {\bibinfo
   {journal} {Phys. Rev. C}\ }\textbf {\bibinfo {volume} {105}},\ \bibinfo
  {pages} {014302} (\bibinfo {year} {2022})}\BibitemShut {NoStop}%
\bibitem [{\citenamefont {Hu}\ \emph {et~al.}(2021)\citenamefont {Hu},
  \citenamefont {Jiang}, \citenamefont {Miyagi}, \citenamefont {Sun},
  \citenamefont {Ekström}, \citenamefont {Forssén}, \citenamefont {Hagen},
  \citenamefont {Holt}, \citenamefont {Papenbrock}, \citenamefont {Stroberg},\
  and\ \citenamefont {Vernon}}]{ref:Hu_22}%
  \BibitemOpen
  \bibfield  {author} {\bibinfo {author} {\bibfnamefont {B.}~\bibnamefont
  {Hu}}, \bibinfo {author} {\bibfnamefont {W.}~\bibnamefont {Jiang}}, \bibinfo
  {author} {\bibfnamefont {T.}~\bibnamefont {Miyagi}}, \bibinfo {author}
  {\bibfnamefont {Z.}~\bibnamefont {Sun}}, \bibinfo {author} {\bibfnamefont
  {A.}~\bibnamefont {Ekström}}, \bibinfo {author} {\bibfnamefont
  {C.}~\bibnamefont {Forssén}}, \bibinfo {author} {\bibfnamefont
  {G.}~\bibnamefont {Hagen}}, \bibinfo {author} {\bibfnamefont {J.~D.}\
  \bibnamefont {Holt}}, \bibinfo {author} {\bibfnamefont {T.}~\bibnamefont
  {Papenbrock}}, \bibinfo {author} {\bibfnamefont {S.~R.}\ \bibnamefont
  {Stroberg}}, \ and\ \bibinfo {author} {\bibfnamefont {I.}~\bibnamefont
  {Vernon}},\ }\href@noop {} {\enquote {\bibinfo {title} {Ab initio predictions
  link the neutron skin of ${}^{208}$pb to nuclear forces},}\ } (\bibinfo
  {year} {2021}),\ \Eprint {http://arxiv.org/abs/2112.01125} {arXiv:2112.01125
  [nucl-th]} \BibitemShut {NoStop}%
\bibitem [{\citenamefont {Stroberg}\ \emph {et~al.}(2021)\citenamefont
  {Stroberg}, \citenamefont {Holt}, \citenamefont {Schwenk},\ and\
  \citenamefont {Simonis}}]{ref:Stroberg_21}%
  \BibitemOpen
  \bibfield  {author} {\bibinfo {author} {\bibfnamefont {S.~R.}\ \bibnamefont
  {Stroberg}}, \bibinfo {author} {\bibfnamefont {J.~D.}\ \bibnamefont {Holt}},
  \bibinfo {author} {\bibfnamefont {A.}~\bibnamefont {Schwenk}}, \ and\
  \bibinfo {author} {\bibfnamefont {J.}~\bibnamefont {Simonis}},\ }\href
  {\doibase 10.1103/PhysRevLett.126.022501} {\bibfield  {journal} {\bibinfo
  {journal} {Phys. Rev. Lett.}\ }\textbf {\bibinfo {volume} {126}},\ \bibinfo
  {pages} {022501} (\bibinfo {year} {2021})}\BibitemShut {NoStop}%
\bibitem [{\citenamefont {Brown}\ \emph {et~al.}(1982)\citenamefont {Brown},
  \citenamefont {Wildenthal}, \citenamefont {Chung}, \citenamefont {Massen},
  \citenamefont {Bernas}, \citenamefont {Bernstein}, \citenamefont {Miskimen},
  \citenamefont {Brown},\ and\ \citenamefont {Madsen}}]{ref:Brown_82}%
  \BibitemOpen
  \bibfield  {author} {\bibinfo {author} {\bibfnamefont {B.~A.}\ \bibnamefont
  {Brown}}, \bibinfo {author} {\bibfnamefont {B.~H.}\ \bibnamefont
  {Wildenthal}}, \bibinfo {author} {\bibfnamefont {W.}~\bibnamefont {Chung}},
  \bibinfo {author} {\bibfnamefont {S.~E.}\ \bibnamefont {Massen}}, \bibinfo
  {author} {\bibfnamefont {M.}~\bibnamefont {Bernas}}, \bibinfo {author}
  {\bibfnamefont {A.~M.}\ \bibnamefont {Bernstein}}, \bibinfo {author}
  {\bibfnamefont {R.}~\bibnamefont {Miskimen}}, \bibinfo {author}
  {\bibfnamefont {V.~R.}\ \bibnamefont {Brown}}, \ and\ \bibinfo {author}
  {\bibfnamefont {V.~A.}\ \bibnamefont {Madsen}},\ }\href
  {https://link.aps.org/doi/10.1103/PhysRevC.26.2247} {\bibfield  {journal}
  {\bibinfo  {journal} {Phys. Rev. C}\ }\textbf {\bibinfo {volume} {26}},\
  \bibinfo {pages} {2247} (\bibinfo {year} {1982})}\BibitemShut {NoStop}%
\bibitem [{\citenamefont {Tsukiyama}\ \emph {et~al.}(2012)\citenamefont
  {Tsukiyama}, \citenamefont {Bogner},\ and\ \citenamefont
  {Schwenk}}]{ref:Tsukiyama_12}%
  \BibitemOpen
  \bibfield  {author} {\bibinfo {author} {\bibfnamefont {K.}~\bibnamefont
  {Tsukiyama}}, \bibinfo {author} {\bibfnamefont {S.~K.}\ \bibnamefont
  {Bogner}}, \ and\ \bibinfo {author} {\bibfnamefont {A.}~\bibnamefont
  {Schwenk}},\ }\href {\doibase 10.1103/PhysRevC.85.061304} {\bibfield
  {journal} {\bibinfo  {journal} {Phys. Rev. C}\ }\textbf {\bibinfo {volume}
  {85}},\ \bibinfo {pages} {061304(R)} (\bibinfo {year} {2012})}\BibitemShut
  {NoStop}%
\bibitem [{\citenamefont {Bogner}\ \emph {et~al.}(2014)\citenamefont {Bogner},
  \citenamefont {Hergert}, \citenamefont {Holt}, \citenamefont {Schwenk},
  \citenamefont {Binder}, \citenamefont {Calci}, \citenamefont {Langhammer},\
  and\ \citenamefont {Roth}}]{ref:Bogner_14}%
  \BibitemOpen
  \bibfield  {author} {\bibinfo {author} {\bibfnamefont {S.~K.}\ \bibnamefont
  {Bogner}}, \bibinfo {author} {\bibfnamefont {H.}~\bibnamefont {Hergert}},
  \bibinfo {author} {\bibfnamefont {J.~D.}\ \bibnamefont {Holt}}, \bibinfo
  {author} {\bibfnamefont {A.}~\bibnamefont {Schwenk}}, \bibinfo {author}
  {\bibfnamefont {S.}~\bibnamefont {Binder}}, \bibinfo {author} {\bibfnamefont
  {A.}~\bibnamefont {Calci}}, \bibinfo {author} {\bibfnamefont
  {J.}~\bibnamefont {Langhammer}}, \ and\ \bibinfo {author} {\bibfnamefont
  {R.}~\bibnamefont {Roth}},\ }\href {\doibase 10.1103/PhysRevLett.113.142501}
  {\bibfield  {journal} {\bibinfo  {journal} {Phys. Rev. Lett.}\ }\textbf
  {\bibinfo {volume} {113}},\ \bibinfo {pages} {142501} (\bibinfo {year}
  {2014})}\BibitemShut {NoStop}%
\bibitem [{\citenamefont {Stroberg}\ \emph {et~al.}(2017)\citenamefont
  {Stroberg}, \citenamefont {Calci}, \citenamefont {Hergert}, \citenamefont
  {Holt}, \citenamefont {Bogner}, \citenamefont {Roth},\ and\ \citenamefont
  {Schwenk}}]{ref:Stroberg_17}%
  \BibitemOpen
  \bibfield  {author} {\bibinfo {author} {\bibfnamefont {S.~R.}\ \bibnamefont
  {Stroberg}}, \bibinfo {author} {\bibfnamefont {A.}~\bibnamefont {Calci}},
  \bibinfo {author} {\bibfnamefont {H.}~\bibnamefont {Hergert}}, \bibinfo
  {author} {\bibfnamefont {J.~D.}\ \bibnamefont {Holt}}, \bibinfo {author}
  {\bibfnamefont {S.~K.}\ \bibnamefont {Bogner}}, \bibinfo {author}
  {\bibfnamefont {R.}~\bibnamefont {Roth}}, \ and\ \bibinfo {author}
  {\bibfnamefont {A.}~\bibnamefont {Schwenk}},\ }\href {\doibase
  10.1103/PhysRevLett.118.032502} {\bibfield  {journal} {\bibinfo  {journal}
  {Phys. Rev. Lett.}\ }\textbf {\bibinfo {volume} {118}},\ \bibinfo {pages}
  {032502} (\bibinfo {year} {2017})}\BibitemShut {NoStop}%
\bibitem [{\citenamefont {Stroberg}\ \emph {et~al.}(2019)\citenamefont
  {Stroberg}, \citenamefont {Bogner}, \citenamefont {Hergert},\ and\
  \citenamefont {Holt}}]{ref:Stroberg_19}%
  \BibitemOpen
  \bibfield  {author} {\bibinfo {author} {\bibfnamefont {S.~R.}\ \bibnamefont
  {Stroberg}}, \bibinfo {author} {\bibfnamefont {S.~K.}\ \bibnamefont
  {Bogner}}, \bibinfo {author} {\bibfnamefont {H.}~\bibnamefont {Hergert}}, \
  and\ \bibinfo {author} {\bibfnamefont {J.~D.}\ \bibnamefont {Holt}},\ }\href
  {\doibase 10.1146/annurev-nucl-101917-021120} {\bibfield  {journal} {\bibinfo
   {journal} {Ann. Rev. Nucl. Part. Sci.}\ }\textbf {\bibinfo {volume} {69}},\
  \bibinfo {pages} {307} (\bibinfo {year} {2019})}\BibitemShut {NoStop}%
\bibitem [{\citenamefont {Parzuchowski}\ \emph {et~al.}(2017)\citenamefont
  {Parzuchowski}, \citenamefont {Stroberg}, \citenamefont {Navr{\'{a}}til},
  \citenamefont {Hergert},\ and\ \citenamefont {Bogner}}]{ref:Parzuchowski_17}%
  \BibitemOpen
  \bibfield  {author} {\bibinfo {author} {\bibfnamefont {N.~M.}\ \bibnamefont
  {Parzuchowski}}, \bibinfo {author} {\bibfnamefont {S.~R.}\ \bibnamefont
  {Stroberg}}, \bibinfo {author} {\bibfnamefont {P.}~\bibnamefont
  {Navr{\'{a}}til}}, \bibinfo {author} {\bibfnamefont {H.}~\bibnamefont
  {Hergert}}, \ and\ \bibinfo {author} {\bibfnamefont {S.~K.}\ \bibnamefont
  {Bogner}},\ }\href {\doibase 10.1103/PhysRevC.96.034324} {\bibfield
  {journal} {\bibinfo  {journal} {Phys. Rev. C}\ }\textbf {\bibinfo {volume}
  {96}},\ \bibinfo {pages} {034324} (\bibinfo {year} {2017})},\ \Eprint
  {http://arxiv.org/abs/1705.05511} {arXiv:1705.05511} \BibitemShut {NoStop}%
\bibitem [{\citenamefont {Hebeler}\ \emph {et~al.}(2011)\citenamefont
  {Hebeler}, \citenamefont {Bogner}, \citenamefont {Furnstahl}, \citenamefont
  {Nogga},\ and\ \citenamefont {Schwenk}}]{ref:Hebeler_11}%
  \BibitemOpen
  \bibfield  {author} {\bibinfo {author} {\bibfnamefont {K.}~\bibnamefont
  {Hebeler}}, \bibinfo {author} {\bibfnamefont {S.~K.}\ \bibnamefont {Bogner}},
  \bibinfo {author} {\bibfnamefont {R.~J.}\ \bibnamefont {Furnstahl}}, \bibinfo
  {author} {\bibfnamefont {A.}~\bibnamefont {Nogga}}, \ and\ \bibinfo {author}
  {\bibfnamefont {A.}~\bibnamefont {Schwenk}},\ }\href@noop {} {\bibfield
  {journal} {\bibinfo  {journal} {Phys. Rev. C}\ }\textbf {\bibinfo {volume}
  {83}},\ \bibinfo {pages} {031301(R)} (\bibinfo {year} {2011})}\BibitemShut
  {NoStop}%
\bibitem [{\citenamefont {Simonis}\ \emph {et~al.}(2016)\citenamefont
  {Simonis}, \citenamefont {Hebeler}, \citenamefont {Holt}, \citenamefont
  {Men\'{e}ndez},\ and\ \citenamefont {Schwenk}}]{ref:Simonis_16}%
  \BibitemOpen
  \bibfield  {author} {\bibinfo {author} {\bibfnamefont {J.}~\bibnamefont
  {Simonis}}, \bibinfo {author} {\bibfnamefont {K.}~\bibnamefont {Hebeler}},
  \bibinfo {author} {\bibfnamefont {J.~D.}\ \bibnamefont {Holt}}, \bibinfo
  {author} {\bibfnamefont {J.}~\bibnamefont {Men\'{e}ndez}}, \ and\ \bibinfo
  {author} {\bibfnamefont {A.}~\bibnamefont {Schwenk}},\ }\href {\doibase
  10.1103/PhysRevC.93.011302} {\bibfield  {journal} {\bibinfo  {journal} {Phys.
  Rev. C}\ }\textbf {\bibinfo {volume} {93}},\ \bibinfo {pages} {011302(R)}
  (\bibinfo {year} {2016})}\BibitemShut {NoStop}%
\bibitem [{\citenamefont {Bogner}\ \emph {et~al.}(2007)\citenamefont {Bogner},
  \citenamefont {Furnstahl},\ and\ \citenamefont {Perry}}]{ref:Bogner_07}%
  \BibitemOpen
  \bibfield  {author} {\bibinfo {author} {\bibfnamefont {S.~K.}\ \bibnamefont
  {Bogner}}, \bibinfo {author} {\bibfnamefont {R.~J.}\ \bibnamefont
  {Furnstahl}}, \ and\ \bibinfo {author} {\bibfnamefont {R.~J.}\ \bibnamefont
  {Perry}},\ }\href@noop {} {\bibfield  {journal} {\bibinfo  {journal} {Phys.
  Rev. C}\ }\textbf {\bibinfo {volume} {75}},\ \bibinfo {pages} {061001(R)}
  (\bibinfo {year} {2007})}\BibitemShut {NoStop}%
\bibitem [{\citenamefont {Entem}\ and\ \citenamefont
  {Machleidt}(2003)}]{ref:Entem_03}%
  \BibitemOpen
  \bibfield  {author} {\bibinfo {author} {\bibfnamefont {D.~R.}\ \bibnamefont
  {Entem}}\ and\ \bibinfo {author} {\bibfnamefont {R.}~\bibnamefont
  {Machleidt}},\ }\href@noop {} {\bibfield  {journal} {\bibinfo  {journal}
  {Phys. Rev. C}\ }\textbf {\bibinfo {volume} {68}},\ \bibinfo {pages}
  {041001(R)} (\bibinfo {year} {2003})}\BibitemShut {NoStop}%
\bibitem [{\citenamefont {Simonis}\ \emph {et~al.}(2017)\citenamefont
  {Simonis}, \citenamefont {Stroberg}, \citenamefont {Hebeler}, \citenamefont
  {Holt},\ and\ \citenamefont {Schwenk}}]{ref:Simonis_17}%
  \BibitemOpen
  \bibfield  {author} {\bibinfo {author} {\bibfnamefont {J.}~\bibnamefont
  {Simonis}}, \bibinfo {author} {\bibfnamefont {S.~R.}\ \bibnamefont
  {Stroberg}}, \bibinfo {author} {\bibfnamefont {K.}~\bibnamefont {Hebeler}},
  \bibinfo {author} {\bibfnamefont {J.~D.}\ \bibnamefont {Holt}}, \ and\
  \bibinfo {author} {\bibfnamefont {A.}~\bibnamefont {Schwenk}},\ }\href@noop
  {} {\bibfield  {journal} {\bibinfo  {journal} {Phys. Rev. C}\ }\textbf
  {\bibinfo {volume} {96}},\ \bibinfo {pages} {014303} (\bibinfo {year}
  {2017})}\BibitemShut {NoStop}%
\bibitem [{\citenamefont {Morris}\ \emph {et~al.}(2018)\citenamefont {Morris},
  \citenamefont {Simonis}, \citenamefont {Stroberg}, \citenamefont {Stumpf},
  \citenamefont {Hagen}, \citenamefont {Holt}, \citenamefont {Jansen},
  \citenamefont {Papenbrock}, \citenamefont {Roth},\ and\ \citenamefont
  {Schwenk}}]{ref:Morris_18}%
  \BibitemOpen
  \bibfield  {author} {\bibinfo {author} {\bibfnamefont {T.~D.}\ \bibnamefont
  {Morris}}, \bibinfo {author} {\bibfnamefont {J.}~\bibnamefont {Simonis}},
  \bibinfo {author} {\bibfnamefont {S.~R.}\ \bibnamefont {Stroberg}}, \bibinfo
  {author} {\bibfnamefont {C.}~\bibnamefont {Stumpf}}, \bibinfo {author}
  {\bibfnamefont {G.}~\bibnamefont {Hagen}}, \bibinfo {author} {\bibfnamefont
  {J.~D.}\ \bibnamefont {Holt}}, \bibinfo {author} {\bibfnamefont {G.~R.}\
  \bibnamefont {Jansen}}, \bibinfo {author} {\bibfnamefont {T.}~\bibnamefont
  {Papenbrock}}, \bibinfo {author} {\bibfnamefont {R.}~\bibnamefont {Roth}}, \
  and\ \bibinfo {author} {\bibfnamefont {A.}~\bibnamefont {Schwenk}},\
  }\href@noop {} {\bibfield  {journal} {\bibinfo  {journal} {Phys. Rev. Lett.}\
  }\textbf {\bibinfo {volume} {120}},\ \bibinfo {pages} {152503} (\bibinfo
  {year} {2018})}\BibitemShut {NoStop}%
\bibitem [{\citenamefont {Brown}\ and\ \citenamefont
  {Richter}(2006)}]{ref:USDB}%
  \BibitemOpen
  \bibfield  {author} {\bibinfo {author} {\bibfnamefont {B.~A.}\ \bibnamefont
  {Brown}}\ and\ \bibinfo {author} {\bibfnamefont {W.~A.}\ \bibnamefont
  {Richter}},\ }\href@noop {} {\bibfield  {journal} {\bibinfo  {journal} {Phys.
  Rev. C}\ }\textbf {\bibinfo {volume} {74}},\ \bibinfo {pages} {034315}
  (\bibinfo {year} {2006})}\BibitemShut {NoStop}%
\bibitem [{\citenamefont {Brown}\ and\ \citenamefont
  {Rae}(2014)}]{ref:NushellX}%
  \BibitemOpen
  \bibfield  {author} {\bibinfo {author} {\bibfnamefont {B.~A.}\ \bibnamefont
  {Brown}}\ and\ \bibinfo {author} {\bibfnamefont {W.~D.~M.}\ \bibnamefont
  {Rae}},\ }\href@noop {} {\bibfield  {journal} {\bibinfo  {journal} {Nucl.
  Data Sheets}\ }\textbf {\bibinfo {volume} {120}},\ \bibinfo {pages} {115}
  (\bibinfo {year} {2014})}\BibitemShut {NoStop}%
\bibitem [{ref({\natexlab{a}})}]{ref:nutbar}%
  \BibitemOpen
  \href {\doibase http://doi.org/10.5281/zenodo.496190} {}\bibinfo
  {howpublished} {\url{https://github.com/ragnarstroberg/nutbar}}
  ({\natexlab{a}})\BibitemShut {NoStop}%
\bibitem [{\citenamefont {Ruotsalainen}\ \emph {et~al.}(2019)\citenamefont
  {Ruotsalainen}, \citenamefont {Henderson}, \citenamefont {Hackman},
  \citenamefont {Sargsyan}, \citenamefont {Launey}, \citenamefont {Saxena},
  \citenamefont {Srivastava}, \citenamefont {Stroberg}, \citenamefont {Grahn},
  \citenamefont {Pakarinen}, \citenamefont {Ball}, \citenamefont {Julin},
  \citenamefont {Greenlees}, \citenamefont {Smallcombe}, \citenamefont
  {Andreoiu}, \citenamefont {Bernier}, \citenamefont {Bowry}, \citenamefont
  {Buckner}, \citenamefont {Caballero-Folch}, \citenamefont {Chester},
  \citenamefont {Cruz}, \citenamefont {Evitts}, \citenamefont {Frederick},
  \citenamefont {Garnsworthy}, \citenamefont {Holl}, \citenamefont {Kurkjian},
  \citenamefont {Kisliuk}, \citenamefont {Leach}, \citenamefont {McGee},
  \citenamefont {Measures}, \citenamefont {M\"ucher}, \citenamefont {Park},
  \citenamefont {Sarazin}, \citenamefont {Smith}, \citenamefont {Southall},
  \citenamefont {Starosta}, \citenamefont {Svensson}, \citenamefont {Whitmore},
  \citenamefont {Williams},\ and\ \citenamefont {Wu}}]{ref:Ruotsalainen_19}%
  \BibitemOpen
  \bibfield  {author} {\bibinfo {author} {\bibfnamefont {P.}~\bibnamefont
  {Ruotsalainen}}, \bibinfo {author} {\bibfnamefont {J.}~\bibnamefont
  {Henderson}}, \bibinfo {author} {\bibfnamefont {G.}~\bibnamefont {Hackman}},
  \bibinfo {author} {\bibfnamefont {G.~H.}\ \bibnamefont {Sargsyan}}, \bibinfo
  {author} {\bibfnamefont {K.~D.}\ \bibnamefont {Launey}}, \bibinfo {author}
  {\bibfnamefont {A.}~\bibnamefont {Saxena}}, \bibinfo {author} {\bibfnamefont
  {P.~C.}\ \bibnamefont {Srivastava}}, \bibinfo {author} {\bibfnamefont
  {S.~R.}\ \bibnamefont {Stroberg}}, \bibinfo {author} {\bibfnamefont
  {T.}~\bibnamefont {Grahn}}, \bibinfo {author} {\bibfnamefont
  {J.}~\bibnamefont {Pakarinen}}, \bibinfo {author} {\bibfnamefont {G.~C.}\
  \bibnamefont {Ball}}, \bibinfo {author} {\bibfnamefont {R.}~\bibnamefont
  {Julin}}, \bibinfo {author} {\bibfnamefont {P.~T.}\ \bibnamefont
  {Greenlees}}, \bibinfo {author} {\bibfnamefont {J.}~\bibnamefont
  {Smallcombe}}, \bibinfo {author} {\bibfnamefont {C.}~\bibnamefont
  {Andreoiu}}, \bibinfo {author} {\bibfnamefont {N.}~\bibnamefont {Bernier}},
  \bibinfo {author} {\bibfnamefont {M.}~\bibnamefont {Bowry}}, \bibinfo
  {author} {\bibfnamefont {M.}~\bibnamefont {Buckner}}, \bibinfo {author}
  {\bibfnamefont {R.}~\bibnamefont {Caballero-Folch}}, \bibinfo {author}
  {\bibfnamefont {A.}~\bibnamefont {Chester}}, \bibinfo {author} {\bibfnamefont
  {S.}~\bibnamefont {Cruz}}, \bibinfo {author} {\bibfnamefont {L.~J.}\
  \bibnamefont {Evitts}}, \bibinfo {author} {\bibfnamefont {R.}~\bibnamefont
  {Frederick}}, \bibinfo {author} {\bibfnamefont {A.~B.}\ \bibnamefont
  {Garnsworthy}}, \bibinfo {author} {\bibfnamefont {M.}~\bibnamefont {Holl}},
  \bibinfo {author} {\bibfnamefont {A.}~\bibnamefont {Kurkjian}}, \bibinfo
  {author} {\bibfnamefont {D.}~\bibnamefont {Kisliuk}}, \bibinfo {author}
  {\bibfnamefont {K.~G.}\ \bibnamefont {Leach}}, \bibinfo {author}
  {\bibfnamefont {E.}~\bibnamefont {McGee}}, \bibinfo {author} {\bibfnamefont
  {J.}~\bibnamefont {Measures}}, \bibinfo {author} {\bibfnamefont
  {D.}~\bibnamefont {M\"ucher}}, \bibinfo {author} {\bibfnamefont
  {J.}~\bibnamefont {Park}}, \bibinfo {author} {\bibfnamefont {F.}~\bibnamefont
  {Sarazin}}, \bibinfo {author} {\bibfnamefont {J.~K.}\ \bibnamefont {Smith}},
  \bibinfo {author} {\bibfnamefont {D.}~\bibnamefont {Southall}}, \bibinfo
  {author} {\bibfnamefont {K.}~\bibnamefont {Starosta}}, \bibinfo {author}
  {\bibfnamefont {C.~E.}\ \bibnamefont {Svensson}}, \bibinfo {author}
  {\bibfnamefont {K.}~\bibnamefont {Whitmore}}, \bibinfo {author}
  {\bibfnamefont {M.}~\bibnamefont {Williams}}, \ and\ \bibinfo {author}
  {\bibfnamefont {C.~Y.}\ \bibnamefont {Wu}},\ }\href@noop {} {\bibfield
  {journal} {\bibinfo  {journal} {Physical Review C}\ }\textbf {\bibinfo
  {volume} {99}},\ \bibinfo {pages} {051301(R)} (\bibinfo {year}
  {2019})}\BibitemShut {NoStop}%
\bibitem [{\citenamefont {NNDC}()}]{ref:ENSDF}%
  \BibitemOpen
  \bibfield  {author} {\bibinfo {author} {\bibnamefont {NNDC}},\ }\href@noop {}
  {\enquote {\bibinfo {title} {{Evaluated Nuclear Structure Data File
  (ENSDF)}},}\ }\BibitemShut {NoStop}%
\bibitem [{\citenamefont {Henderson}\ \emph {et~al.}(2022)\citenamefont
  {Henderson} \emph {et~al.}}]{ref:Henderson_21}%
  \BibitemOpen
  \bibfield  {author} {\bibinfo {author} {\bibfnamefont {J.}~\bibnamefont
  {Henderson}} \emph {et~al.},\ }\href@noop {} {\bibfield  {journal} {\bibinfo
  {journal} {Submitted to Physical Review C}\ } (\bibinfo {year}
  {2022})}\BibitemShut {NoStop}%
\bibitem [{\citenamefont {Petkov}\ \emph {et~al.}(2017)\citenamefont {Petkov},
  \citenamefont {M\"uller-Gatermann}, \citenamefont {Dewald}, \citenamefont
  {Blazhev}, \citenamefont {Fransen}, \citenamefont {Jolie}, \citenamefont
  {Scholz}, \citenamefont {Zell},\ and\ \citenamefont
  {Zilges}}]{ref:Petkov_17}%
  \BibitemOpen
  \bibfield  {author} {\bibinfo {author} {\bibfnamefont {P.}~\bibnamefont
  {Petkov}}, \bibinfo {author} {\bibfnamefont {C.}~\bibnamefont
  {M\"uller-Gatermann}}, \bibinfo {author} {\bibfnamefont {A.}~\bibnamefont
  {Dewald}}, \bibinfo {author} {\bibfnamefont {A.}~\bibnamefont {Blazhev}},
  \bibinfo {author} {\bibfnamefont {C.}~\bibnamefont {Fransen}}, \bibinfo
  {author} {\bibfnamefont {J.}~\bibnamefont {Jolie}}, \bibinfo {author}
  {\bibfnamefont {P.}~\bibnamefont {Scholz}}, \bibinfo {author} {\bibfnamefont
  {K.~O.}\ \bibnamefont {Zell}}, \ and\ \bibinfo {author} {\bibfnamefont
  {A.}~\bibnamefont {Zilges}},\ }\href@noop {} {\bibfield  {journal} {\bibinfo
  {journal} {Phys. Rev. C}\ }\textbf {\bibinfo {volume} {96}},\ \bibinfo
  {pages} {034326} (\bibinfo {year} {2017})}\BibitemShut {NoStop}%
\bibitem [{\citenamefont {VonMoss}\ \emph {et~al.}(2015)\citenamefont
  {VonMoss}, \citenamefont {Tabor}, \citenamefont {Tripathi}, \citenamefont
  {Volya}, \citenamefont {Abromeit}, \citenamefont {Bender}, \citenamefont
  {Caussyn}, \citenamefont {Dungan}, \citenamefont {Kravvaris}, \citenamefont
  {Kuchera}, \citenamefont {Lubna}, \citenamefont {Miller}, \citenamefont
  {Parker},\ and\ \citenamefont {Tai}}]{ref:VonMoss_15}%
  \BibitemOpen
  \bibfield  {author} {\bibinfo {author} {\bibfnamefont {J.~M.}\ \bibnamefont
  {VonMoss}}, \bibinfo {author} {\bibfnamefont {S.~L.}\ \bibnamefont {Tabor}},
  \bibinfo {author} {\bibfnamefont {V.}~\bibnamefont {Tripathi}}, \bibinfo
  {author} {\bibfnamefont {A.}~\bibnamefont {Volya}}, \bibinfo {author}
  {\bibfnamefont {B.}~\bibnamefont {Abromeit}}, \bibinfo {author}
  {\bibfnamefont {P.~C.}\ \bibnamefont {Bender}}, \bibinfo {author}
  {\bibfnamefont {D.~D.}\ \bibnamefont {Caussyn}}, \bibinfo {author}
  {\bibfnamefont {R.}~\bibnamefont {Dungan}}, \bibinfo {author} {\bibfnamefont
  {K.}~\bibnamefont {Kravvaris}}, \bibinfo {author} {\bibfnamefont {M.~P.}\
  \bibnamefont {Kuchera}}, \bibinfo {author} {\bibfnamefont {R.}~\bibnamefont
  {Lubna}}, \bibinfo {author} {\bibfnamefont {S.}~\bibnamefont {Miller}},
  \bibinfo {author} {\bibfnamefont {J.~J.}\ \bibnamefont {Parker}}, \ and\
  \bibinfo {author} {\bibfnamefont {P.-L.}\ \bibnamefont {Tai}},\ }\href
  {https://link.aps.org/doi/10.1103/PhysRevC.92.034301} {\bibfield  {journal}
  {\bibinfo  {journal} {Phys. Rev. C}\ }\textbf {\bibinfo {volume} {92}},\
  \bibinfo {pages} {034301} (\bibinfo {year} {2015})}\BibitemShut {NoStop}%
\bibitem [{\citenamefont {Wendt}\ \emph {et~al.}()\citenamefont {Wendt},
  \citenamefont {Taprogge}, \citenamefont {Reiter}, \citenamefont {Golubev},
  \citenamefont {Grawe}, \citenamefont {Pietri}, \citenamefont {Boutachkov},
  \citenamefont {Algora}, \citenamefont {Ameil}, \citenamefont {Bentley},
  \citenamefont {Blazhev}, \citenamefont {Bloor}, \citenamefont {Nara~Singh},
  \citenamefont {Bowry}, \citenamefont {Bracco}, \citenamefont {Braun},
  \citenamefont {Camera}, \citenamefont {Cederk\"all}, \citenamefont {Crespi},
  \citenamefont {de~la Salle}, \citenamefont {DiJulio}, \citenamefont
  {Doornenbal}, \citenamefont {Geibel}, \citenamefont {Gellanki}, \citenamefont
  {Gerl}, \citenamefont {Gr\ifmmode~\mbox{\c{e}}\else \c{e}\fi{}bosz},
  \citenamefont {Guastalla}, \citenamefont {Habermann}, \citenamefont
  {Hackstein}, \citenamefont {Hoischen}, \citenamefont {Jungclaus},
  \citenamefont {Merch\'an}, \citenamefont {Million}, \citenamefont {Morales},
  \citenamefont {Moschner}, \citenamefont {Podoly\'ak}, \citenamefont
  {Pietralla}, \citenamefont {Ralet}, \citenamefont {Reese}, \citenamefont
  {Rudolph}, \citenamefont {Scruton}, \citenamefont {Siebeck}, \citenamefont
  {Warr}, \citenamefont {Wieland},\ and\ \citenamefont
  {Wollersheim}}]{ref:Wendt_14}%
  \BibitemOpen
  \bibfield  {author} {\bibinfo {author} {\bibfnamefont {A.}~\bibnamefont
  {Wendt}}, \bibinfo {author} {\bibfnamefont {J.}~\bibnamefont {Taprogge}},
  \bibinfo {author} {\bibfnamefont {P.}~\bibnamefont {Reiter}}, \bibinfo
  {author} {\bibfnamefont {P.}~\bibnamefont {Golubev}}, \bibinfo {author}
  {\bibfnamefont {H.}~\bibnamefont {Grawe}}, \bibinfo {author} {\bibfnamefont
  {S.}~\bibnamefont {Pietri}}, \bibinfo {author} {\bibfnamefont
  {P.}~\bibnamefont {Boutachkov}}, \bibinfo {author} {\bibfnamefont
  {A.}~\bibnamefont {Algora}}, \bibinfo {author} {\bibfnamefont
  {F.}~\bibnamefont {Ameil}}, \bibinfo {author} {\bibfnamefont {M.~A.}\
  \bibnamefont {Bentley}}, \bibinfo {author} {\bibfnamefont {A.}~\bibnamefont
  {Blazhev}}, \bibinfo {author} {\bibfnamefont {D.}~\bibnamefont {Bloor}},
  \bibinfo {author} {\bibfnamefont {B.~S.}\ \bibnamefont {Nara~Singh}},
  \bibinfo {author} {\bibfnamefont {M.}~\bibnamefont {Bowry}}, \bibinfo
  {author} {\bibfnamefont {A.}~\bibnamefont {Bracco}}, \bibinfo {author}
  {\bibfnamefont {N.}~\bibnamefont {Braun}}, \bibinfo {author} {\bibfnamefont
  {F.}~\bibnamefont {Camera}}, \bibinfo {author} {\bibfnamefont
  {J.}~\bibnamefont {Cederk\"all}}, \bibinfo {author} {\bibfnamefont
  {F.}~\bibnamefont {Crespi}}, \bibinfo {author} {\bibfnamefont
  {A.}~\bibnamefont {de~la Salle}}, \bibinfo {author} {\bibfnamefont
  {D.}~\bibnamefont {DiJulio}}, \bibinfo {author} {\bibfnamefont
  {P.}~\bibnamefont {Doornenbal}}, \bibinfo {author} {\bibfnamefont
  {K.}~\bibnamefont {Geibel}}, \bibinfo {author} {\bibfnamefont
  {J.}~\bibnamefont {Gellanki}}, \bibinfo {author} {\bibfnamefont
  {J.}~\bibnamefont {Gerl}}, \bibinfo {author} {\bibfnamefont {J.}~\bibnamefont
  {Gr\ifmmode~\mbox{\c{e}}\else \c{e}\fi{}bosz}}, \bibinfo {author}
  {\bibfnamefont {G.}~\bibnamefont {Guastalla}}, \bibinfo {author}
  {\bibfnamefont {T.}~\bibnamefont {Habermann}}, \bibinfo {author}
  {\bibfnamefont {M.}~\bibnamefont {Hackstein}}, \bibinfo {author}
  {\bibfnamefont {R.}~\bibnamefont {Hoischen}}, \bibinfo {author}
  {\bibfnamefont {A.}~\bibnamefont {Jungclaus}}, \bibinfo {author}
  {\bibfnamefont {E.}~\bibnamefont {Merch\'an}}, \bibinfo {author}
  {\bibfnamefont {B.}~\bibnamefont {Million}}, \bibinfo {author} {\bibfnamefont
  {A.}~\bibnamefont {Morales}}, \bibinfo {author} {\bibfnamefont
  {K.}~\bibnamefont {Moschner}}, \bibinfo {author} {\bibfnamefont
  {Z.}~\bibnamefont {Podoly\'ak}}, \bibinfo {author} {\bibfnamefont
  {N.}~\bibnamefont {Pietralla}}, \bibinfo {author} {\bibfnamefont
  {D.}~\bibnamefont {Ralet}}, \bibinfo {author} {\bibfnamefont
  {M.}~\bibnamefont {Reese}}, \bibinfo {author} {\bibfnamefont
  {D.}~\bibnamefont {Rudolph}}, \bibinfo {author} {\bibfnamefont
  {L.}~\bibnamefont {Scruton}}, \bibinfo {author} {\bibfnamefont
  {B.}~\bibnamefont {Siebeck}}, \bibinfo {author} {\bibfnamefont
  {N.}~\bibnamefont {Warr}}, \bibinfo {author} {\bibfnamefont {O.}~\bibnamefont
  {Wieland}}, \ and\ \bibinfo {author} {\bibfnamefont {H.~J.}\ \bibnamefont
  {Wollersheim}},\ }\href@noop {} {\bibfield  {journal} {\bibinfo  {journal}
  {Phys. Rev. C}\ }\textbf {\bibinfo {volume} {90}},\ \bibinfo {pages}
  {054301}}\BibitemShut {NoStop}%
\bibitem [{\citenamefont {Siegel}\ and\ \citenamefont
  {Zamick}(1970)}]{ref:Siegel_70}%
  \BibitemOpen
  \bibfield  {author} {\bibinfo {author} {\bibfnamefont {S.}~\bibnamefont
  {Siegel}}\ and\ \bibinfo {author} {\bibfnamefont {L.}~\bibnamefont
  {Zamick}},\ }\href@noop {} {\bibfield  {journal} {\bibinfo  {journal}
  {Nuclear Physics A}\ }\textbf {\bibinfo {volume} {145}},\ \bibinfo {pages}
  {89} (\bibinfo {year} {1970})}\BibitemShut {NoStop}%
\bibitem [{\citenamefont {Scherpenzeel}\ \emph {et~al.}(1980)\citenamefont
  {Scherpenzeel}, \citenamefont {Engelbertink}, \citenamefont {Aarts},
  \citenamefont {{van der Poel}},\ and\ \citenamefont
  {Arciszewski}}]{ref:Scherpenzeel_80}%
  \BibitemOpen
  \bibfield  {author} {\bibinfo {author} {\bibfnamefont {D.}~\bibnamefont
  {Scherpenzeel}}, \bibinfo {author} {\bibfnamefont {G.}~\bibnamefont
  {Engelbertink}}, \bibinfo {author} {\bibfnamefont {H.}~\bibnamefont {Aarts}},
  \bibinfo {author} {\bibfnamefont {C.}~\bibnamefont {{van der Poel}}}, \ and\
  \bibinfo {author} {\bibfnamefont {H.}~\bibnamefont {Arciszewski}},\ }\href
  {\doibase https://doi.org/10.1016/0375-9474(80)90304-8} {\bibfield  {journal}
  {\bibinfo  {journal} {Nuclear Physics A}\ }\textbf {\bibinfo {volume}
  {349}},\ \bibinfo {pages} {513} (\bibinfo {year} {1980})}\BibitemShut
  {NoStop}%
\bibitem [{\citenamefont {{Okano}}\ \emph {et~al.}(1960)\citenamefont
  {{Okano}}, \citenamefont {{Tabata}},\ and\ \citenamefont
  {{Fukuda}}}]{ref:Okano_60}%
  \BibitemOpen
  \bibfield  {author} {\bibinfo {author} {\bibfnamefont {K.}~\bibnamefont
  {{Okano}}}, \bibinfo {author} {\bibfnamefont {T.}~\bibnamefont {{Tabata}}}, \
  and\ \bibinfo {author} {\bibfnamefont {K.}~\bibnamefont {{Fukuda}}},\
  }\href@noop {} {\bibfield  {journal} {\bibinfo  {journal} {J.Phys.Soc.Japan}\
  }\textbf {\bibinfo {volume} {15}},\ \bibinfo {pages} {1556} (\bibinfo {year}
  {1960})}\BibitemShut {NoStop}%
\bibitem [{\citenamefont {Byrski}\ \emph {et~al.}(1974)\citenamefont {Byrski},
  \citenamefont {Beck}, \citenamefont {Engelstein}, \citenamefont {Forterre},\
  and\ \citenamefont {Knipper}}]{ref:Byrski_74}%
  \BibitemOpen
  \bibfield  {author} {\bibinfo {author} {\bibfnamefont {T.}~\bibnamefont
  {Byrski}}, \bibinfo {author} {\bibfnamefont {F.}~\bibnamefont {Beck}},
  \bibinfo {author} {\bibfnamefont {P.}~\bibnamefont {Engelstein}}, \bibinfo
  {author} {\bibfnamefont {M.}~\bibnamefont {Forterre}}, \ and\ \bibinfo
  {author} {\bibfnamefont {A.}~\bibnamefont {Knipper}},\ }\href {\doibase
  https://doi.org/10.1016/0375-9474(74)90281-4} {\bibfield  {journal} {\bibinfo
   {journal} {Nuclear Physics A}\ }\textbf {\bibinfo {volume} {223}},\ \bibinfo
  {pages} {125} (\bibinfo {year} {1974})}\BibitemShut {NoStop}%
\bibitem [{\citenamefont {Bohr}\ and\ \citenamefont
  {Mottelson}()}]{ref:BohrMottelson}%
  \BibitemOpen
  \bibfield  {author} {\bibinfo {author} {\bibfnamefont {A.}~\bibnamefont
  {Bohr}}\ and\ \bibinfo {author} {\bibfnamefont {B.~R.}\ \bibnamefont
  {Mottelson}},\ }\href@noop {} {\emph {\bibinfo {title} {Nuclear Structure
  Vol. II: Nuclear Deformations}}}\BibitemShut {NoStop}%
\bibitem [{\citenamefont {Shimizu}\ \emph {et~al.}(2019)\citenamefont
  {Shimizu}, \citenamefont {Mizusaki}, \citenamefont {Utsuno},\ and\
  \citenamefont {Tsunoda}}]{Shimizu2019}%
  \BibitemOpen
  \bibfield  {author} {\bibinfo {author} {\bibfnamefont {N.}~\bibnamefont
  {Shimizu}}, \bibinfo {author} {\bibfnamefont {T.}~\bibnamefont {Mizusaki}},
  \bibinfo {author} {\bibfnamefont {Y.}~\bibnamefont {Utsuno}}, \ and\ \bibinfo
  {author} {\bibfnamefont {Y.}~\bibnamefont {Tsunoda}},\ }\href {\doibase
  10.1016/j.cpc.2019.06.011} {\bibfield  {journal} {\bibinfo  {journal}
  {Comput. Phys. Commun.}\ }\textbf {\bibinfo {volume} {244}},\ \bibinfo
  {pages} {372} (\bibinfo {year} {2019})}\BibitemShut {NoStop}%
\bibitem [{\citenamefont {Warburton}\ and\ \citenamefont
  {Brown}(1992)}]{ref:Warburton_92PRC}%
  \BibitemOpen
  \bibfield  {author} {\bibinfo {author} {\bibfnamefont {E.~K.}\ \bibnamefont
  {Warburton}}\ and\ \bibinfo {author} {\bibfnamefont {B.~A.}\ \bibnamefont
  {Brown}},\ }\href@noop {} {\bibfield  {journal} {\bibinfo  {journal}
  {Physical Review C}\ }\textbf {\bibinfo {volume} {46}},\ \bibinfo {pages}
  {923} (\bibinfo {year} {1992})}\BibitemShut {NoStop}%
\bibitem [{\citenamefont {Warburton}\ \emph {et~al.}(1992)\citenamefont
  {Warburton}, \citenamefont {Brown},\ and\ \citenamefont
  {Millener}}]{ref:Warburton_92}%
  \BibitemOpen
  \bibfield  {author} {\bibinfo {author} {\bibfnamefont {E.~K.}\ \bibnamefont
  {Warburton}}, \bibinfo {author} {\bibfnamefont {B.~A.}\ \bibnamefont
  {Brown}}, \ and\ \bibinfo {author} {\bibfnamefont {D.~J.}\ \bibnamefont
  {Millener}},\ }\href@noop {} {\bibfield  {journal} {\bibinfo  {journal}
  {Physics Letters B}\ }\textbf {\bibinfo {volume} {293}},\ \bibinfo {pages}
  {7} (\bibinfo {year} {1992})}\BibitemShut {NoStop}%
\bibitem [{\citenamefont {Stanton}\ and\ \citenamefont
  {Bartlett}(1993)}]{ref:Stanton_93}%
  \BibitemOpen
  \bibfield  {author} {\bibinfo {author} {\bibfnamefont {J.~F.}\ \bibnamefont
  {Stanton}}\ and\ \bibinfo {author} {\bibfnamefont {R.~J.}\ \bibnamefont
  {Bartlett}},\ }\href {\doibase 10.1063/1.464746} {\bibfield  {journal}
  {\bibinfo  {journal} {J. Chem. Phys.}\ }\textbf {\bibinfo {volume} {98}},\
  \bibinfo {pages} {7029} (\bibinfo {year} {1993})}\BibitemShut {NoStop}%
\bibitem [{\citenamefont {Lee}\ \emph {et~al.}(1984)\citenamefont {Lee},
  \citenamefont {Kucharski},\ and\ \citenamefont {Bartlett}}]{ref:Lee_84}%
  \BibitemOpen
  \bibfield  {author} {\bibinfo {author} {\bibfnamefont {Y.~S.}\ \bibnamefont
  {Lee}}, \bibinfo {author} {\bibfnamefont {S.~A.}\ \bibnamefont {Kucharski}},
  \ and\ \bibinfo {author} {\bibfnamefont {R.~J.}\ \bibnamefont {Bartlett}},\
  }\href@noop {} {\bibfield  {journal} {\bibinfo  {journal} {Journal of
  Chemical Physics}\ }\textbf {\bibinfo {volume} {81}},\ \bibinfo {pages}
  {5906} (\bibinfo {year} {1984})}\BibitemShut {NoStop}%
\bibitem [{\citenamefont {Watts}\ and\ \citenamefont
  {Bartlett}(1996)}]{ref:Watts_96}%
  \BibitemOpen
  \bibfield  {author} {\bibinfo {author} {\bibfnamefont {J.~D.}\ \bibnamefont
  {Watts}}\ and\ \bibinfo {author} {\bibfnamefont {R.~J.}\ \bibnamefont
  {Bartlett}},\ }\href@noop {} {\bibfield  {journal} {\bibinfo  {journal}
  {Chemical Physics Letters}\ }\textbf {\bibinfo {volume} {258}},\ \bibinfo
  {pages} {581} (\bibinfo {year} {1996})}\BibitemShut {NoStop}%
\bibitem [{ref({\natexlab{b}})}]{ref:imsrgcode}%
  \BibitemOpen
  \href {https://github.com/ragnarstroberg/imsrg} {}\bibinfo {howpublished}
  {\url{https://github.com/ragnarstroberg/imsrg}} ({\natexlab{b}})\BibitemShut
  {NoStop}%
\bibitem [{\citenamefont {Sanderson}(2010)}]{ref:Armadillo}%
  \BibitemOpen
  \bibfield  {author} {\bibinfo {author} {\bibfnamefont {C.}~\bibnamefont
  {Sanderson}},\ }\href {http://arma.sourceforge.net/armadillo_nicta_2010.pdf}
  {\bibfield  {journal} {\bibinfo  {journal} {Techical Report, NICTA}\ }
  (\bibinfo {year} {2010})}\BibitemShut {NoStop}%
\bibitem [{\citenamefont {Ring}\ and\ \citenamefont
  {Schuck}(1980)}]{ref:Ring_80}%
  \BibitemOpen
  \bibfield  {author} {\bibinfo {author} {\bibfnamefont {P.}~\bibnamefont
  {Ring}}\ and\ \bibinfo {author} {\bibfnamefont {P.}~\bibnamefont {Schuck}},\
  }\href@noop {} {\emph {\bibinfo {title} {{The Nuclear Many-Body Problem}}}},\
  \bibinfo {edition} {1st}\ ed.\ (\bibinfo  {publisher} {Springer-Verlag},\
  \bibinfo {address} {Berlin, Heidelberg},\ \bibinfo {year} {1980})\BibitemShut
  {NoStop}%
\end{thebibliography}%

\end{document}